\pdfoutput=1
\documentclass[12pt]{article}

\usepackage[margin=1in]{geometry}
\usepackage[T1]{fontenc}
\usepackage[utf8]{inputenc}
\usepackage{mathptmx}
\usepackage{microtype}
\usepackage{amsmath}
\usepackage{bm}
\usepackage{graphicx}
\usepackage[superscript]{cite}

\usepackage{booktabs}
\usepackage{caption}
\usepackage{setspace}
\usepackage{authblk}
\usepackage{xcolor}
\usepackage{enumitem}
\usepackage{ragged2e}
\usepackage{hyperref}
\ifdefined\pdfminorversion\pdfminorversion=7\fi
\newcommand{\qtyunit}[2]{\ensuremath{#1\,\mathrm{#2}}}

\newcommand{\Qz}{\mathbf{Q}_0}
\newcommand{\Dplus}{D^{+}}
\newcommand{\dd}{\boldsymbol{\delta}}
\newcommand{\kk}{\mathbf{k}}
\newcommand{\PP}{\mathbf{P}}
\newcommand{\QQ}{\mathbf{Q}}
\newcommand{\RR}{\mathbf{R}}
\newcommand{\vq}{\mathbf{q}}

\hypersetup{
  colorlinks=true,
  citecolor=blue,
  linkcolor=blue,
  urlcolor=blue,
  pdftitle={Long-lived light-driven bimagnons at finite momentum in NiO},
  pdfauthor={Xian P. Yang et al.}
}

\title{
Long-lived light-driven bimagnons at finite momentum in NiO
}

\author[1]{Xian P. Yang}
\author[1]{Christopher T. Parzyck}
\author[2]{Martin Cross}
\author[1,3]{Ella Blake}
\author[1]{Octave Duros}
\author[1]{Yinchuan Lv}
\author[4]{Hari Paudyal}
\author[5]{Che-Min Lin}
\author[1]{Sharon S. Philip}
\author[2]{Lingjia Shen}
\author[2]{Daniel Jost}
\author[2]{Giacomo Coslovich}
\author[1,2]{Joshua J. Turner}
\author[2]{Choongwon Seo}
\author[2]{Elyse Schriber}
\author[4,6]{Michael E. Flatt\'e}
\author[7]{Xiaoqin Li}
\author[1,8,9]{Yuri Suzuki}
\author[4]{Durga Paudyal}
\author[1,8]{Harold Y. Hwang}
\author[2*]{Georgi L. Dakovski}
\author[2*]{Matthias C. Hoffmann}
\author[1*]{Wei-Sheng Lee}
\author[2*]{Alexander H. Reid}

\affil[1]{Stanford Institute for Materials and Energy Sciences, SLAC National Accelerator Laboratory, Menlo Park, California 94025, USA}
\affil[2]{Linac Coherent Light Source, SLAC National Accelerator Laboratory, Menlo Park, California 94025, USA}
\affil[3]{Department of Physics, Stanford University, Stanford, California 94305, USA}
\affil[4]{Department of Physics and Astronomy, University of Iowa, Iowa City, Iowa 52242, USA}
\affil[5]{Department of Materials Science and Engineering, National Taiwan University of Science and Technology (NTUST), Taipei, Taiwan}
\affil[6]{Department of Applied Physics and Science Education, Eindhoven University of Technology, Eindhoven 5612 AZ, The Netherlands}
\affil[7]{Department of Physics and Center for Complex Quantum Systems, The University of Texas at Austin, Austin, Texas 78712, USA}
\affil[8]{Department of Applied Physics, Stanford University, Stanford, California 94305, USA}
\affil[9]{Geballe Laboratory for Advanced Materials, Stanford University, Stanford, California 94305, USA}
\affil[*]{Corresponding authors.}

\date{}

\begin{document}

\maketitle

\setstretch{1.15}

{\setstretch{1.15}\bfseries

Light can manipulate magnetic order and its excitations on ultrafast timescales\cite{refBeaurepaire1996,refKimel2005,refKirilyuk2010,refKampfrath2011,refNemec2018,refLeenders2024,refRadovskaia2026,refKimel2026}, including through resonantly driven phonons\cite{refDisa2020,refAfanasiev2021}. In antiferromagnets, light can also create correlated magnon pairs, or bimagnons\cite{ref3,ref6,refZhao2004,refBossini2016,refSchoenfeld2025}. Unlike single magnons, whose damping has been measured and modeled in detail\cite{refBayrakci2013,refZhitomirsky2013,refChoe2025}, little is known about the lifetime, interactions and decay of such correlated pairs: prior two-magnon measurements show broad linewidths\cite{ref1,refGhiringhelli2009,refBetto2017} or coherences that dephase within 200~fs\cite{refBossini2016,refBossini2019}. Here we pump the phonon-assisted two-magnon absorption of NiO with mid-infrared pulses, far below its bandgap\cite{ref8,ref1,ref6}, and probe the resulting magnetic dynamics at finite momentum with time-resolved resonant inelastic X-ray scattering (RIXS)\cite{refMitrano2020,refCao2019}. Upon photoexcitation, the one-magnon linewidth and energy are essentially unchanged, indicating that the long-range magnetic order is preserved without the photocarrier-induced demagnetization characteristic of above-bandgap excitation. A pump-induced two-magnon signal appears on the energy-gain (anti-Stokes) side of the spectrum. Its spectral weight peaks 4~ps after the pump, well beyond our 0.5~ps time resolution, indicative of a delayed growth likely due to momentum redistribution of the light-induced magnon pairs across the Brillouin zone\cite{refBayrakci2013,refHua2026,refHua2024}. This two-magnon signal decays with a time constant of $28\pm11$~ps, exceeding the previously reported two-magnon dephasing time by more than two orders of magnitude. These results establish long-lived light-driven bimagnon correlations beyond the one-magnon state, akin to the higher-order correlations measured in ultracold gases\cite{refGring2012,refSchweigler2017}.

\par}

\section*{Main}

Antiferromagnets hold great technological potential for spin-based microelectronics. Their spin excitations have intrinsically high frequencies, and they neither produce significant stray fields nor require external bias fields for operation. Optical control of antiferromagnets is a promising route to such technologies\cite{refKimel2005,refKampfrath2011,refNemec2018,refRadovskaia2026,refKimel2026}. However, the transduction of energy from light into spin excitations at finite momentum remains a challenge. An optical photon carries almost no momentum on the scale of the Brillouin zone, so light excites single magnons only near the zone center\cite{refKampfrath2011}. This limits the addressable frequency range and access to propagating magnon modes that exist away from the zone center. One way around this limit is to couple light to multi-magnon excitations. For a multi-magnon excitation, only the total momentum is fixed, so each individual magnon can have a short wavelength anywhere in the Brillouin zone.

Multi-magnon excitations also depend on the interaction between magnons. Spin-wave theory is the standard description of magnetic excitations. In its linear form it treats magnons as independent particles, and the interaction between magnons is added only as a higher-order correction\cite{refZhitomirsky2013,refBayrakci2013}. This is justified near equilibrium, where few magnons are excited, but a driven magnet need not satisfy this condition. Magnon pairs, the lowest-order multi-magnon excitation, are especially sensitive to these interactions. Light modulates
the exchange coupling $J_{ij}$ of a bond and so couples to
$\mathbf S_i\cdot\mathbf S_j$, which creates or annihilates two magnons on
neighboring spins, a bimagnon\cite{ref3,ref6}. The two magnons of such a pair are correlated, and the two-magnon response reflects this correlation. After photoexcitation, the two-magnon
response therefore depends on whether the excited magnons are correlated in
pairs as well as on how many there are. It can also relax on a different
timescale from the one-magnon occupations. 

Prior two-magnon measurements all report fast dynamics. In KNiF$_3$, femtosecond pulses drive magnon pairs impulsively through
stimulated Raman scattering\cite{refBossini2016,refBossini2019}, and the coherent two-magnon oscillations dephase within 200~fs\cite{refBossini2016}. The equilibrium
two-magnon linewidth of NiO corresponds to sub-picosecond times in both Raman
scattering\cite{ref1} and resonant inelastic X-ray scattering
(RIXS)\cite{refBetto2017}. Photoexcitation across the bandgap points the same
way. The two-magnon Raman signal of YBa$_2$Cu$_3$O$_{6.1}$ shows
demagnetization within 90~fs\cite{refYang2020}, and in Sr$_2$IrO$_4$ the
in-plane magnetic correlations recover within a few
picoseconds\cite{refDean2016}. These timescales record the loss of phase or the recovery of order, so they
invite the expectation that driven pairs are short-lived. 

That expectation has not been directly tested experimentally. A short dephasing time sets no upper
limit on how long the two-magnon response of driven pairs persists. The existing above-bandgap experiments,
moreover, create photocarriers, which suppress long-range magnetic order and
shape the transient magnetic correlations\cite{refDean2016,refMazzone2021}. In
YBa$_2$Cu$_3$O$_{6.1}$, for example, the magnetic response is governed by the
slow relaxation of the photoexcited carriers\cite{refYang2020}. Measuring how long the two-magnon response of driven pairs persists without photocarriers therefore requires a pump energy below all electronic excitations. In addition, it requires a probe that can resolve the excitation momentum, because optical probes see only pairs with near zero total momentum\cite{refZhao2004,refYang2020,refBossini2016}.

RIXS, which records the energy and momentum that scattered X-rays leave in the sample, resolves two-magnon excitations in both\cite{refHill2008}. The advent of high-brightness X-ray free-electron lasers has now enabled RIXS probes to be combined with above-bandgap optical excitation to study spin dynamics\cite{refDean2016,refMazzone2021,ref10}, offering new access to multi-magnons. In this work, we instead pair time-resolved RIXS with a below-bandgap resonant pump in NiO, a prototypical antiferromagnet. This combination gives access to one-magnon and two-magnon dynamics at finite momentum without photocarriers.

The pump we applied is resonant with the phonon-assisted two-magnon absorption near 0.25~eV, which lies far below the lowest electronic
excitations, including the electric-dipole forbidden $d$-$d$ transitions near 1.1~eV and the charge-transfer gap\cite{ref1,ref8}. It therefore creates bimagnons without photocarriers that would disrupt the magnetic order or damp the pumped excitations. On the energy-gain (anti-Stokes) side of the RIXS spectrum, the X-rays gain energy by annihilating excitations
already present, so the pump-induced anti-Stokes weight rises and falls with the pumped population, revealing a long-lived light-induced bimagnon correlation.

\section*{Finite-momentum two-magnon spectroscopy}

\begin{figure}[tbp]
\centering
\includegraphics[width=\linewidth]{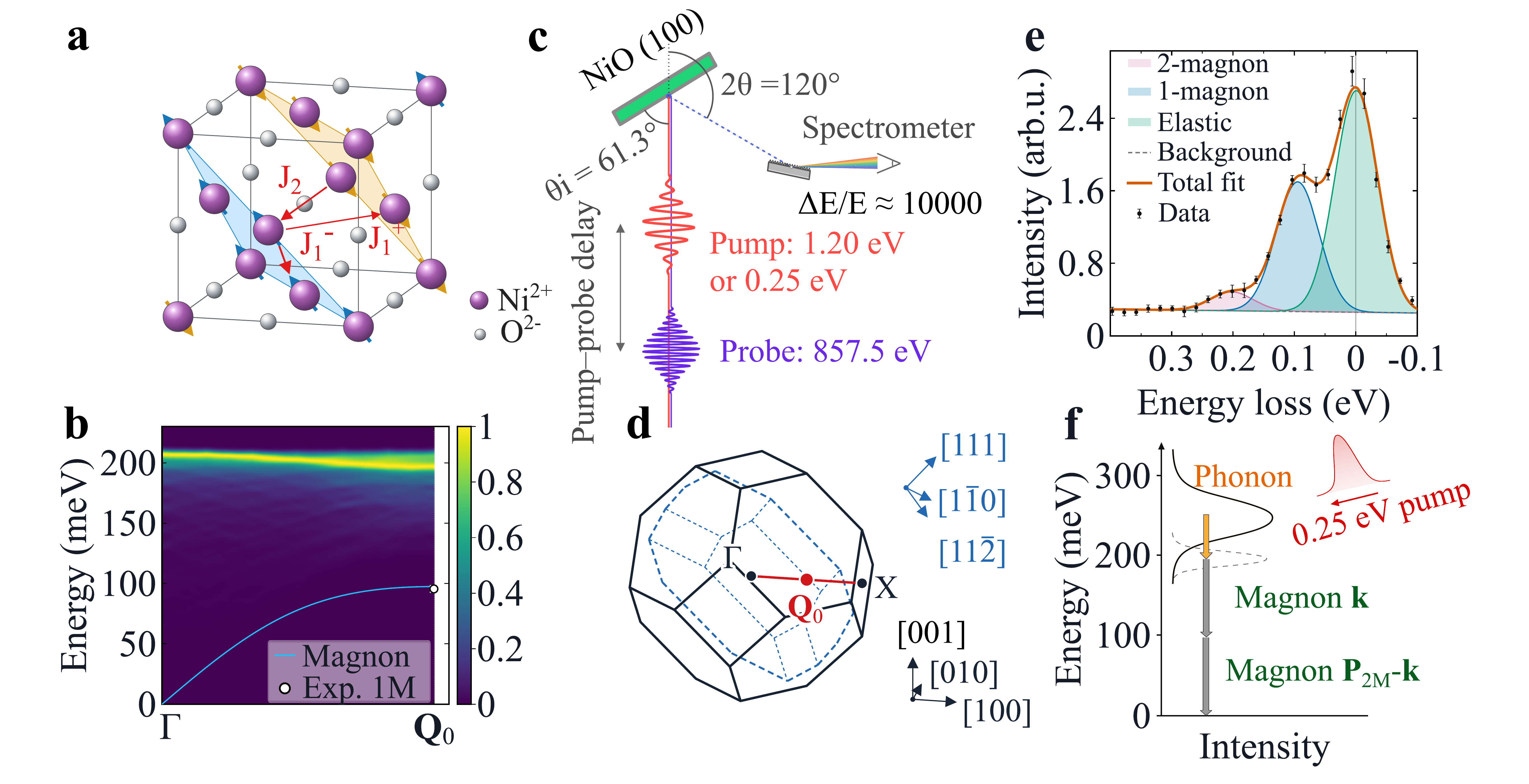}
\caption{
\textbf{Resonant bimagnon excitation and finite-momentum two-magnon detection.}
\textbf{a}, Type-II antiferromagnetic order in NiO and the exchange couplings $J_1$ and $J_2$.
\textbf{b}, Calculated one-magnon dispersion (cyan) and bare two-magnon density
of states along [100] from $\Gamma$ to $\Qz$, obtained from linear spin-wave
theory with the full first-principles exchange network (Methods;
Supplementary Note~1). The color scale shows the two-magnon density of states,
normalized at each total momentum. The white circle marks the experimental
one-magnon energy at $\Qz$, $95.5\pm2.7$~meV, from the equilibrium fit in \textbf{e}.
\textbf{c}, Pump-probe Ni $L_3$-edge RIXS geometry for a NiO (100) film, with scattering angle $2\theta=120^\circ$ and X-ray incidence angle $61.3^\circ$.
\textbf{d}, Structural (fcc) Brillouin zone of NiO (solid) and magnetic Brillouin
zone of the type-II order (dashed blue). The red dot marks the probed momentum
$\Qz=(\tfrac12,0,0)\,2\pi/a$ along the film normal [100], and the black dot marks
the structural zone boundary along [100] at $X=(1,0,0)\,2\pi/a$. Blue arrows represent
[111], the stacking direction of the ferromagnetic (111) planes, and the in-plane
directions [1$\bar1$0] and [11$\bar2$]; black arrows are the cubic axes.
\textbf{e}, Equilibrium RIXS spectrum at $\Qz$ with the one-magnon peak and high-energy two-magnon shoulder. The spectrum is the average of six aligned pre-pump runs, with error bars denoting standard errors across these runs; the fitted components are from a single fit to this average (Methods).
\textbf{f}, Phonon-assisted two-magnon absorption near the 0.25~eV pump energy. The absorbed photon energy is shared between the magnetic excitation and an assisting phonon. The momentum selection rule $\mathbf P_{2M}+\mathbf q_{\rm phonon}=\mathbf k_{\rm photon}$ permits both near-zero and finite total pair momenta, with weights determined by the optical matrix elements.
}
\label{fig:fig1}
\end{figure}

NiO is a type-II antiferromagnet in which the Ni$^{2+}$ moments align
ferromagnetically within (111) planes but antiferromagnetically between
neighboring planes~\cite{refHutchings1972} (Fig.~\ref{fig:fig1}a). Our
scattering geometry corresponds to a momentum transfer along the sample surface normal close
to $\Qz=(\tfrac12,0,0)$ (in units of $2\pi/a$; Fig.~\ref{fig:fig1}c,d).
Figure~\ref{fig:fig1}b shows the magnetic excitations along [100] from
$\Gamma$ to $\Qz$, obtained from a first-principles
calculation that keeps the full exchange network (Supplementary Note~1). At
$\Qz$ the one-magnon branch lies close to its maximum, and the bare two-magnon density of states peaks near \qtyunit{0.2}{eV}. The equilibrium RIXS spectrum at
$\Qz$ (Fig.~\ref{fig:fig1}e) is consistent with this picture. Besides the
elastic line at zero energy loss, it shows a one-magnon (1M) peak near 100~meV and
a broad two-magnon (2M) shoulder near 200~meV that overlaps the calculated two-magnon density
of states, as in previous RIXS
measurements~\cite{refGhiringhelli2009,refBetto2017,ref11}.

Two kinds of excitation contribute to this 2M shoulder at the Ni $L_3$ edge. A \textit{bimagnon} is an intersite excitation created or annihilated by $\mathbf S_i\cdot\mathbf S_j$ on neighboring spins, with $\Delta M_S=0$~\cite{ref6,ref7}, and carries the correlations between bond exchange energies we want to follow. A \textit{double magnon} is a double spin flip on a single Ni ion, with $\Delta M_S=\pm2$~\cite{refGhiringhelli2009}, which the exchange interaction between the core hole and the valence electrons allows in the RIXS intermediate state~\cite{ref11}. The two appear at similar energies, and in NiO double magnons can dominate the equilibrium 2M signal~\cite{ref11}, so the Stokes-side 2M shoulder is not a direct measure of bimagnons. The pump, by contrast, selectively excites only bimagnons. We tune the pump to the phonon-assisted two-magnon absorption near
\qtyunit{0.25}{eV} (Fig.~\ref{fig:fig1}f)~\cite{ref8,ref1}, whose energy
matches the sum of an optical phonon~\cite{ref8,ref15} and the two-magnon Raman
peak~\cite{ref1}. The phonon modulates the superexchange
$\mathbf S_i\cdot\mathbf S_j$ while breaking the inversion symmetry of the
Ni-O-Ni bond, which gives the magnetic excitation an electric-dipole moment and
makes it optically allowed~\cite{ref6,ref7}. Since $\mathbf S_i\cdot\mathbf S_j$ conserves the total $S^z$, the pump creates bimagnons with $\Delta M_S=0$.

The assisting phonon also sets the momentum of the magnon pair. Momentum conservation
requires
$\mathbf P_{2\mathrm M}+\mathbf q_{\mathrm{phonon}}=\mathbf k_{\mathrm{photon}}$,
where $\mathbf P_{2\mathrm M}$ is the total momentum of the magnon pair.
Because $\mathbf k_{\mathrm{photon}}$ is negligible on the scale of the
Brillouin zone, $\mathbf P_{2\mathrm M}\simeq-\mathbf q_{\mathrm{phonon}}$, so
a phonon with finite momentum creates a pair with finite total momentum.
Early analyses of the phonon-assisted two-magnon absorption of NiO considered nearly
zone-center optical phonons~\cite{ref1}, whereas later theories include
phonons from the whole zone and predict a strongly momentum-dependent
absorption~\cite{ref6,ref7}. Time-resolved RIXS probing the
2M response at a defined non-zero momentum allows us to test if the infrared pump excites a
distribution of phonon-magnon configurations rather than only pairs at near-zero momentum.

\section*{Resonant access to the driven magnetic state}

\begin{figure}[tbp]
\centering
\includegraphics[width=\linewidth]{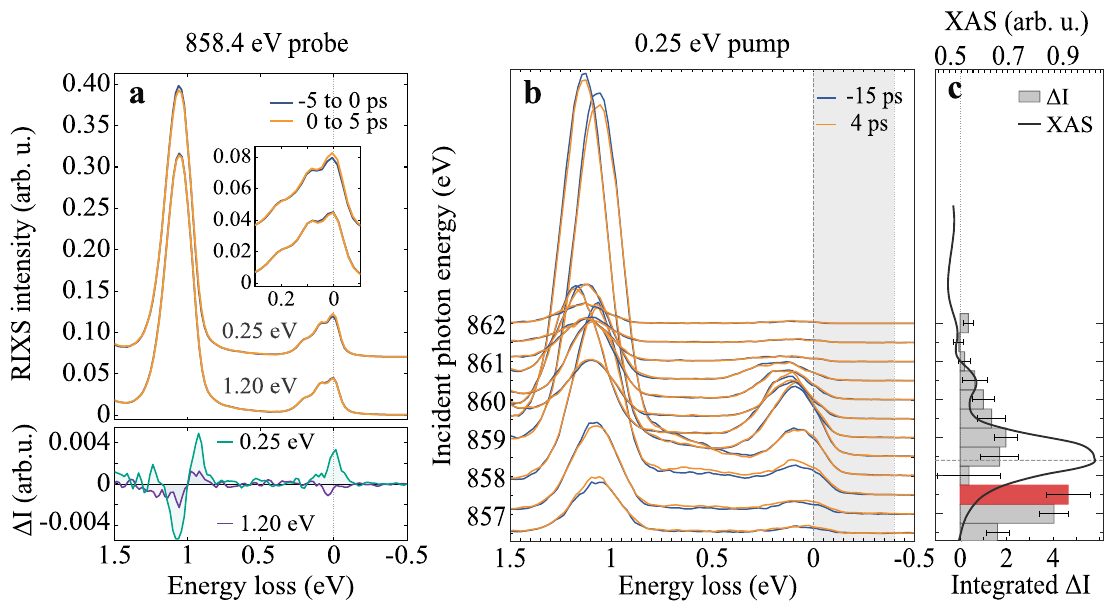}
\caption{
\textbf{Pump-wavelength and probe-energy dependence of the transient response.}
\textbf{a}, RIXS spectra before and after time zero for 0.25~eV and 1.20~eV excitation at an incident X-ray energy of 858.4 eV, with the pump-induced differences below. The two
pairs of spectra are offset vertically. Inset, the same spectra near the
elastic line.  The resonant mid-infrared pump produces the larger anti-Stokes magnetic response despite the stronger optical absorption at 1.20~eV.
\textbf{b}, Spectra measured across the Ni $L_3$ edge before excitation and at \qtyunit{4}{ps} after the 0.25~eV pump. The grey area marks the anti-Stokes
window from 0 to $-0.4$~eV that is integrated in \textbf{c}.
\textbf{c}, Pump-induced weight in this window (bars), compared with the
equilibrium X-ray absorption spectrum (line, top axis). The dashed line marks
the absorption maximum at 858.4~eV. The transient response is enhanced on the low-energy side of the absorption maximum. Error bars are standard errors across runs.  The red bar marks 857.5 eV, which is the incident energy used in the time-dependent measurements in Fig.~\ref{fig:fig3}.
}
\label{fig:fig2}
\end{figure}

We first verify that the pump couples resonantly to the magnetic system by
comparing the transients at $\Qz$ produced by 0.25~eV and 1.20~eV
pulses, with the probe tuned to the Ni $L_3$ edge (Fig.~\ref{fig:fig2}a). The 0.25~eV pump is resonant with the NiO phonon-assisted two-magnon absorption, while the 1.20~eV pump falls within the lowest-lying $d$-$d$ absorption band near 1.1~eV. A
Beer-Lambert estimate based on the low-temperature absorption coefficients of
NiO shows that, in the 120~nm film, the 1.20~eV pulse deposits about 3.5 times
more energy than the 0.25~eV pulse (Supplementary Note~2). Yet the
0.25~eV pump produces a much larger low-energy anti-Stokes
response. The size of the transient is thus determined by the absorption channel, as expected for resonant excitation of the
phonon-assisted two-magnon absorption. The extended anti-Stokes spectra also show no trace of the charge-transfer exciton seen under above-bandgap
excitation~\cite{ref10} (Supplementary Fig.~\ref{fig:sfig3}).

While the pumped
bimagnons ($\Delta M_S=0$) have a similar energy to single-site double spin
flips ($\Delta M_S=\pm2$)~\cite{ref11}, they could be distinguished by
their resonance behaviors as a function of incident photon energy (Fig.~\ref{fig:fig2}b,c). In equilibrium, the resonant profiles of 1M and 2M weights are in qualitative agreement with Nag
\textit{et al.}~\cite{ref11} (Supplementary Fig.~\ref{fig:sfig5}). As a comparison, the pump-induced anti-Stokes weight peaks 0.9~eV below the Ni $L_3$ absorption maximum so it
differs from both the X-ray absorption spectrum (XAS) and the equilibrium
Stokes 2M signal. The contrast is sharpest at \qtyunit{857.5}{eV}, where the
dynamics in Fig.~\ref{fig:fig3} were measured. There the equilibrium Stokes 2M
signal is strongly suppressed, whereas the pump-induced anti-Stokes 2M signal
is strong. Whatever the equilibrium 2M signal consists of, the transient resonates at a different incident energy and so comes mainly from a different excitation. Together with the pump-wavelength selectivity, this
contrast supports assigning the transient to intersite bimagnons rather than
to single-site double magnons.

\section*{Long-lived finite-momentum bimagnon response}

\begin{figure}[tbp]
\centering
\includegraphics[width=\linewidth]{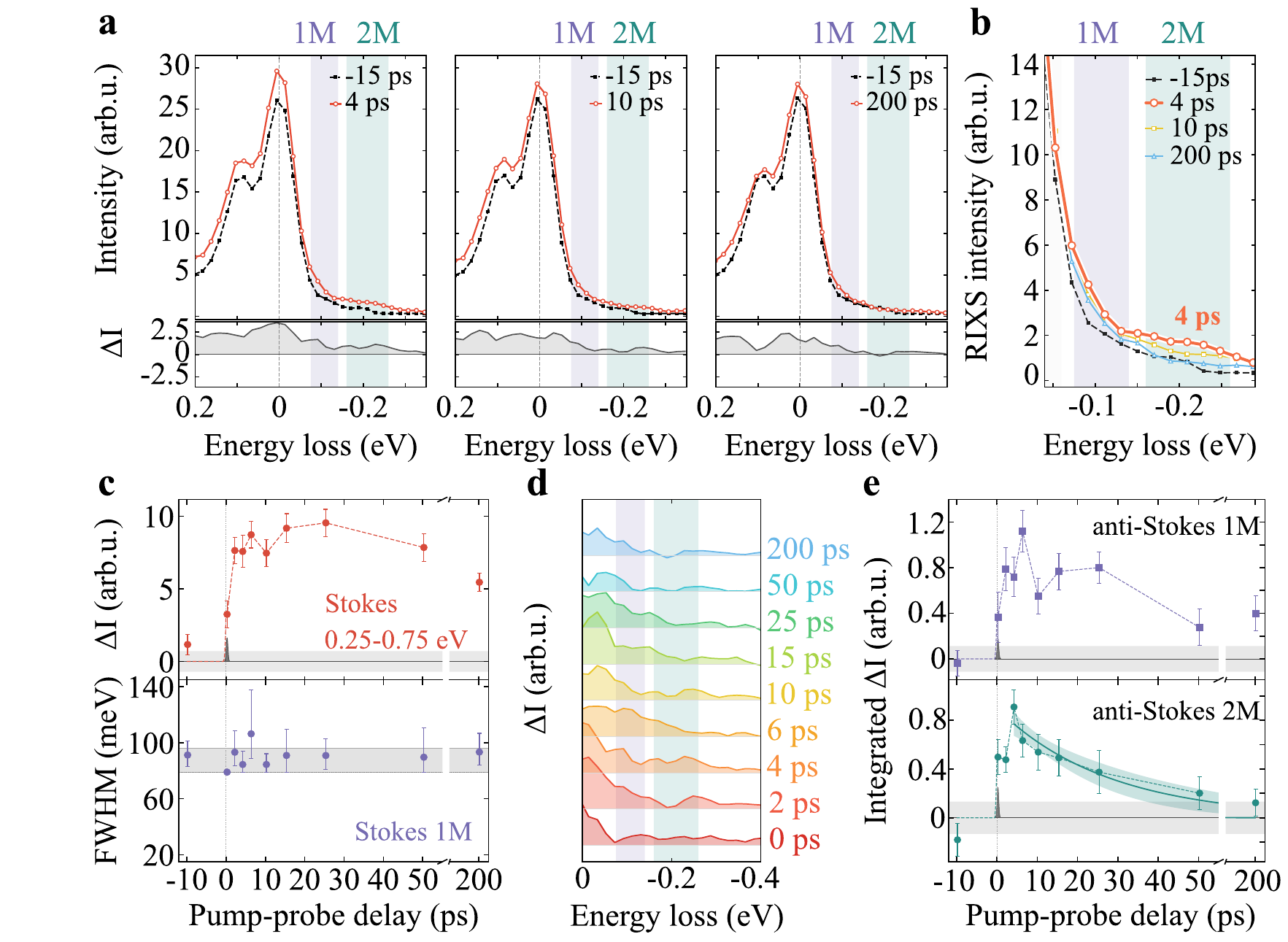}
\caption{
\textbf{Long-lived finite-momentum two-magnon response at $\Qz$.}
\textbf{a}, RIXS spectra at 4, 10 and 200~ps, each compared with the $-15$~ps
reference of the same runs. The resulting difference spectra are shown below;
shading marks the one-magnon (1M, $-140$ to $-75$~meV) and two-magnon (2M,
$-260$ to $-160$~meV) integration windows.
\textbf{b}, Anti-Stokes spectra at 4, 10 and 200~ps together with the same
$-15$~ps reference.
\textbf{c}, Top: pump-induced change integrated over the Stokes-side window
from 0.25 to 0.75~eV, between the elastic/magnon response and the $dd$
excitations; this broad background includes possible thermal contributions.
The gray band is the standard error of the $-10$~ps control. Bottom: measured
full width at half maximum of the Stokes-side 1M peak, with the instrumental
resolution fixed at 79~meV (Methods). Each delay is compared with
the pre-pump reference of the same runs. The gray band spans the outer limits
of the 95\% ranges of all matched pre-pump references. 
\textbf{d}, Difference spectra on the anti-Stokes side from 0 to 200~ps, each relative to the $-15$~ps reference of the same run, vertically offset for clarity.
\textbf{e}, Integrated transient intensity in the anti-Stokes 1M and 2M
windows; the gray band is the standard error
of the $-10$~ps control. The 2M response reaches its maximum at 4~ps and
remains observable beyond 10~ps. The solid line is a weighted
single-exponential fit to the 2M trace from 4 to 200~ps, with its long-delay
offset fixed to zero, yielding $\tau_{\rm 2M}=28\pm11$~ps ($1\sigma$
covariance error; shaded band). The 1M trace reaches its maximum at 6~ps. 
In \textbf{c} and \textbf{e}, points are plotted at the phase-cavity-corrected
delays, 0.22-0.38~ps later than nominal (Methods,
Supplementary Fig.~\ref{fig:sfig_pcav}), and the error bar of each point is
$\pm1$ standard error across runs (95\% run-level bootstrap range for the 1M
width). The grey peak at $t=0$ is a Gaussian with the 0.50~ps FWHM time
resolution of the 0~ps point, centered on that point. Dashed lines are guides
to the eye; around $t=0$ they follow a step broadened by the same resolution.
}
\label{fig:fig3}
\end{figure}

On the energy-gain side, the scattered X-rays gain energy by annihilating
excitations already present in the driven sample. The pump-induced anti-Stokes signal therefore
reflects the population of magnetic excitations that remain after time-zero. Figure~\ref{fig:fig3} shows the time evolution of this signal up to
\qtyunit{200}{ps}. At \qtyunit{4}{ps} the spectrum exhibits a structured
energy-gain profile with substantial weight in the 2M region, alongside a 1M
response at lower energy (Fig.~\ref{fig:fig3}a,b). We quantify the two by tracking the spectral weight within $-260$ to $-160$~meV for the 2M response and from
$-140$ to $-75$~meV for the 1M response. Difference spectra, each taken between the spectrum at a given delay and the
pre-pump spectrum of the same run (Methods), show how the anti-Stokes weight
builds up before it decays (Fig.~\ref{fig:fig3}d).

The 2M signal is largest at \qtyunit{4}{ps} and remains visible beyond
\qtyunit{10}{ps} (Fig.~\ref{fig:fig3}d,e). A single-exponential fit from 4 to
200~ps, with the long-delay offset fixed to zero, gives
$\tau_{\rm 2M}=28\pm11$~ps, much longer than the buildup time of 4~ps. This is also far
slower than the timescales of earlier two-magnon measurements. In KNiF$_3$, driven impulsively by femtosecond pulses, coherent two-magnon
oscillations dephase in $167\pm4$~fs\cite{refBossini2016}. The equilibrium two-magnon linewidth of NiO
corresponds to about 270~fs in Raman scattering\cite{ref1} and to about 28~fs
in RIXS\cite{refBetto2017}. These values describe the loss of coherence or the
spectral broadening, whereas $\tau_{\rm 2M}$ describes how long the driven
finite-momentum 2M response takes to recover.

The Stokes side of the same spectra evolves differently. First, the high-energy background spectral weight 
between 0.25 eV and the $dd$ excitations (Fig.~\ref{fig:fig3}c,
top) rises within 2~ps, keeps growing until 25~ps and still holds about
60\% of its maximum at 200~ps, likely due to thermalization of the system. The 2M weight has fallen to the noise level by 200~ps
(Fig.~\ref{fig:fig3}e), so the 2M signal relaxes on its own timescale,
well before this background.

Second, fitting the Stokes-side 1M peak at every delay and comparing it with
the pre-pump reference of the same runs (Methods), we find no significant
pump-induced broadening at any delay (Fig.~\ref{fig:fig3}c, bottom), unlike the
magnetic excitations seen under above-bandgap
photoexcitation\cite{refMazzone2021,refJost2026}. This
indicates that the resonant pumping does not disrupt the magnetic order and
shows no sign of demagnetization.

The anti-Stokes 1M response also evolves differently from the 2M response. It peaks at
\qtyunit{6}{ps}, after the 2M maximum, and between 4 and 6~ps the 1M weight
still rises while the 2M weight already falls (Fig.~\ref{fig:fig3}e). The
balance between the two magnetic responses thus shifts while the 2M signal
persists. Finally, the delayed 2M maximum shows that the weight at the probed
momentum keeps evolving after the initial optical excitation. This motivates
the redistribution picture discussed next.

\section*{A microscopic route to momentum redistribution}

\begin{figure}[tbp]
\centering
\includegraphics[width=\linewidth]{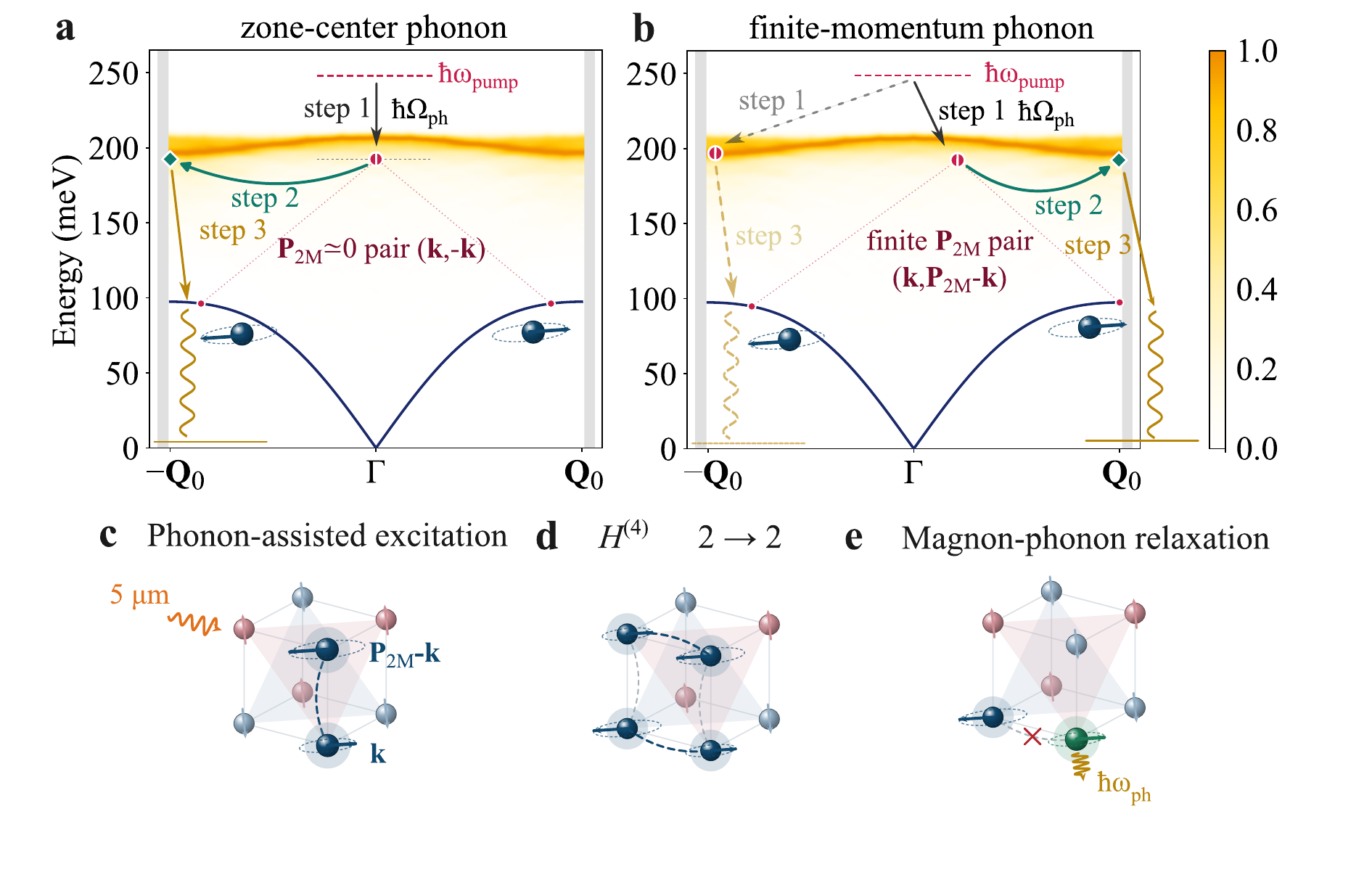}
\caption{
\textbf{A microscopic picture of momentum redistribution.}
\textbf{a,b}, Schematic limiting optical-excitation scenarios on the calculated magnetic excitation spectrum. Yellow indicates the two-magnon continuum from the first-principles
calculation (Fig.~\ref{fig:fig1}b) and blue is the one-magnon dispersion. The red dotted line marks the pump photon energy, $\hbar\omega_{\rm pump}$. The black arrow denotes the assisting phonon energy $\hbar\Omega_{\rm ph}$.
\textbf{a}, A near-zone-center assisting phonon permits two-magnon excitation which has total momentum near $\mathbf{P}_{2\mathrm{M}}= 0$ (step one). It contains magnons with energies and momenta shown by the red dots at the one-magnon dispersion. Subsequent scattering can redistribute weight towards finite pair momenta (green arrow as step two). The yellow pathway near $\Qz$ indicates magnon-phonon relaxation (step three). 
\textbf{b}, Finite-momentum phonons allow direct injection into finite-$P$ momenta (step one), including near $\Qz$ (dashed arrow). Green trajectory indicates possible subsequent redistribution (step two) and the yellow pathway near $\Qz$ indicates magnon-phonon relaxation (step three). These schematics illustrate possible initial distributions and scattering pathways. An initially under-weighted $\Qz$ can gain weight through redistribution in either picture.
\textbf{c}, Phonon-assisted resonant excitation creates magnon pairs in NiO.
\textbf{d}, A representative cross-pair $H^{(4)}$ $2\rightarrow2$ collision. Faint dotted lines indicate the original pair associations and bold dashed lines indicate possible outgoing associations. The outgoing magnons can form pairs with different total momenta while conserving the total energy and crystal momentum.
\textbf{e}, Magnon-phonon scattering exchanges energy and momentum with the lattice and can relax the probed two-magnon response.
}
\label{fig:fig4}
\end{figure}

At our time resolution of 0.50-0.64~ps FWHM (grey peak in
Fig.~\ref{fig:fig3}c,e; Methods), a response that rose with the pump would
reach its full height within about 1~ps, well before the 4~ps maximum. Therefore, the delayed 2M maximum implies that spectral weight continues to transfer to the
probed momentum after the excitation. One possible reason lies in the initial
distribution of pair momenta, which phonon-assisted absorption need not make
uniform. Two limiting cases illustrate the possible starting points
(Fig.~\ref{fig:fig4}a,b). A near-zone-center assisting phonon creates magnon pairs
with total momentum $\mathbf P_{2\mathrm M}\simeq0$, whereas finite-momentum
phonons create pairs directly at finite $\mathbf P_{2\mathrm M}$. Sizeable 1M
and 2M responses are already present at zero delay (Fig.~\ref{fig:fig3}e), so
the probed momentum already acquires weight during the excitation and gains more afterwards. The assisting phonon energy inferred from the experiment,
$\hbar\omega_{\rm pump}-E_{\rm 2M}\simeq\qtyunit{50}{meV}$, lies within the
optical-phonon range of NiO\cite{ref8,ref15}, which makes finite-momentum
phonon-assisted excitation energetically possible.

A mechanism for the subsequent redistribution is the quartic magnon
interaction $H^{(4)}$, whose number-conserving part scatters two magnons into two new magnons\cite{refBayrakci2013,refRomaguera2026,refHua2026,refHua2024}. Resonant soft-X-ray scattering has recently imaged this process in yttrium iron garnet at gigahertz frequencies, where four-magnon scattering spreads microwave-driven magnons over all in-plane propagation directions\cite{refWittrock2026}. Similarly, in NiO, when the colliding magnons come from different optically created pairs, the
outgoing magnons can form new pairs with different total momenta
(Fig.~\ref{fig:fig4}a,b; Supplementary Fig.~\ref{fig:sfig2}). Such cross-pair
collisions move two-magnon weight between momenta while conserving the total
energy and crystal momentum.

To estimate how often these collisions occur, we evaluate the
dilute-limit $2\rightarrow2$ collision rate with the $H^{(4)}$ matrix elements
of the minimal $J_1$-$J_2$ model (Supplementary Note~2). The collision
rate per magnon is $1/\tau_{2\rightarrow2}=K\bar n_{\rm mag}$, where
$\bar n_{\rm mag}$ is the number of pumped magnons per Ni and
$K\simeq410~\mathrm{ps}^{-1}$, varying by less than 10\% with the assumed
pair-momentum distribution. The incident fluence of 34~mJ\,cm$^{-2}$ and
the low-temperature absorption coefficient at 0.25~eV lead to
$\bar n_{\rm mag}\simeq1.24\times10^{-3}\eta$, where $\eta$ is the fraction
of absorbed photons that create a magnon pair. The resulting
collision time, $\tau_{2\rightarrow2}\simeq1.97/\eta$~ps, is a few picoseconds
for order-unity $\eta$, comparable to the 4~ps at which the 2M signal peaks at $\Qz$.

Fig.~\ref{fig:fig4}c-e summarizes the resulting picture.
Phonon-assisted excitation first creates a nonuniform distribution of magnon
pairs. Number-conserving collisions then move weight between momenta without
removing magnetic excitations, which can increase the response at $\Qz$ if
this momentum was initially under-populated. At later times, magnon scattering
together with magnon-phonon coupling moves weight away from the probed
momentum, while energy and momentum flow to the lattice\cite{refMaldonado2019}.

\section*{Discussion}

Our results demonstrate that resonant two-magnon excitations offer a compelling route to selectively couple optical energy into spin excitations, even at finite momenta away from the zone center, corresponding to nanometer length scales. Because the optical excitation is tuned to the phonon-assisted two-magnon absorption far below the bandgap, the absorbed energy goes into the coupled lattice-spin system without creating charge carriers. Intriguingly, the resonantly excited lattice-spin system shows no sign of degradation of the long-range spin order. Together, these conditions provide a favorable environment for hosting long-lived magnetic excitations.

These long-lived excitations, in turn, offer a unique window into the dynamics of bimagnon correlations in NiO. The resonantly excited bimagnon correlations persist more than two orders of magnitude longer than the previously reported two-magnon dephasing time\cite{refBossini2016}. This indicates that the conventional expectation, that two-magnon excitations decay almost immediately into single magnons, does not necessarily hold. The 2M weight at $\Qz$ keeps growing until 4~ps, likely because cross-pair magnon-magnon collisions form new pairs at different total momenta. These findings underscore the importance of magnon-magnon interactions, beyond linear spin-wave theory, in describing the magnetic excitations of an ultrafast light-driven magnet out of equilibrium.

Looking ahead, achieving full control of light-magnon transduction will require a deeper understanding of how two-magnon excitations evolve and interact. We expect the pump fluence to tune the magnon-magnon collision rate, because this rate scales with the pumped magnon density (Supplementary Note~2). It may also be possible to form quasi-equilibrium distributions of two-magnon states, similar to the prethermalized states of isolated quantum systems\cite{refGring2012,refMallayya2019}. More broadly, resonant excitation of coupled lattice-spin systems establishes a promising route toward momentum-selective control of magnetic excitations, opening new opportunities for ultrafast magnonic and spintronic technologies.

\section*{Methods}

\paragraph{Sample growth and characterization.}
NiO films with the cubic rock-salt structure were grown on MgO (100) substrates by pulsed laser deposition. A first layer approximately \qtyunit{60}{nm} thick was deposited at \qtyunit{700}{^{\circ}C} under an oxygen partial pressure of 10~mTorr. The film was then annealed in the same chamber at \qtyunit{700}{^{\circ}C} under an oxygen partial pressure of 250~mTorr for 15~min. The oxygen partial pressure was subsequently returned to the growth condition, and a second approximately \qtyunit{60}{nm} layer was deposited, followed by a second annealing step under the same conditions. The nominal total film thickness was therefore approximately \qtyunit{120}{nm}.

The NiO film was characterized at room temperature by Raman spectroscopy
and X-ray diffraction (XRD), as shown in Supplementary
Fig.~\ref{fig:sfig_growth}. The Raman spectrum shows a two-magnon band
near 1475~cm$^{-1}$, consistent with the room-temperature two-magnon
Raman response reported for NiO\cite{ref1}. The XRD pattern resolves
the MgO (200) substrate and NiO (200) film reflections.

\paragraph{Time-resolved RIXS.}
Time-resolved resonant inelastic X-ray scattering (trRIXS) measurements were performed at the qRIXS endstation of the Linac Coherent Light Source II (LCLS-II), SLAC National Accelerator Laboratory. The sample was held at \qtyunit{25}{K} for the duration of the RIXS measurement.

The magnetic excitations were measured at the Ni $L_3$ edge using soft X-ray pulses at a repetition rate of 33 kHz. The scattering angle was fixed at $2\theta = 120^\circ$, and the X-ray incidence angle was $\theta = 61.3^\circ$. At 857.5~eV this gives a momentum transfer of 0.753~\AA$^{-1}$ close to $\Qz=(\tfrac12,0,0)$
(Fig.~\ref{fig:fig1}c,d). The incidence angle is offset by $1.3^\circ$ from
the specular condition ($\theta = 60^\circ$) to avoid the strong elastic signal at
specular reflection. Unless stated otherwise, the time-dependent data were acquired at an incident photon energy of \qtyunit{857.5}{eV}. For the incident-energy scan in Fig.~\ref{fig:fig2}b,c, the photon energy was varied across the Ni $L_3$ absorption maximum near \qtyunit{858.4}{eV}. The combined energy resolution of the beamline and spectrometer was approximately \qtyunit{79}{meV} measured on carbon tape. The incident X-rays were horizontally polarized, with a pulse duration of about 200 fs and a focal spot of approximately 20~$\mu$m $\times$ 100~$\mu$m (FWHMs). 

The sample was excited with mid-infrared pulses centered at 0.25~eV (wavelength \qtyunit{5}{\mu m}), near the low-temperature phonon-assisted two-magnon absorption peak. The pulses had a nominal duration of 250~fs, were vertically polarized and were collinear with the X-ray beam. The pump spot size at the sample was approximately 74~$\mu$m $\times$ 91~$\mu$m (FWHMs). The incident fluence was 34~mJ\,cm$^{-2}$. Control measurements with 1.20~eV excitation (wavelength \qtyunit{1030}{nm}) used 350 fs, vertically polarized pulses, which had a $0.5^\circ$ offset from the X-ray. The spot size at the sample was approximately 205~$\mu$m $\times$ 201~$\mu$m (FWHMs). The incident fluence was 17.9~mJ\,cm$^{-2}$.

Pump-probe delays ranged from \qtyunit{-15}{ps} to \qtyunit{200}{ps}. Temporal overlap between the optical pump and X-ray probe was established
in two steps. First, the 0.25~eV pulse was cross-correlated
with the 1.20~eV pulse in a 0.5-mm-thick (110)-oriented ZnTe
crystal positioned at the sample location. Sum- and difference-frequency
mixing generated sidebands near 854 and 1297~nm, which were isolated from
the 1.20~eV fundamental using a crossed polarizer placed before a photodiode. Scanning the
relative optical delay yielded a cross-correlation with a full width at
half maximum of approximately 500~fs. Deconvolution with the
$\sim350$-fs 1.20~eV pulse gives an upper bound of
$\sim350$~fs for the mid-infrared pulse duration; additional broadening is
expected from group-velocity mismatch and the finite phase-matching
bandwidth of the ZnTe crystal.

With the relative timing between the two optical pulses fixed, the laser
pulses were then synchronized to the X-rays using an upstream delay stage
common to both pump beams. Time zero was determined from the
X-ray-induced transient reflectivity of a 1.20~eV probe from an
InGaAs-coated Si wafer placed at the sample position. This calibration was repeated at the beginning of each shift and after optical
realignment. Time-zero drifts of up to $\sim$5~ps between shifts were corrected by
re-zeroing the delay stage. Drifts during dataset~(i) were monitored with the
phase-cavity monitor,
and the effective pump-probe time resolution was 0.50-0.64~ps FWHM.

\paragraph{X-ray absorption spectroscopy.}
Static Ni $L_{2,3}$-edge X-ray absorption spectra were measured at Beamline 13-3 of the Stanford Synchrotron Radiation Lightsource (SSRL), SLAC National Accelerator Laboratory. Measurements were performed at \qtyunit{14}{K} in total-electron-yield mode using $\sigma$-polarized X rays. The spectra were normalized to the incident X-ray flux. The Ni $L_3$ spectrum shown in Fig.~\ref{fig:fig2}c provides the equilibrium reference for the incident-energy dependence of the transient RIXS response.

\paragraph{Datasets.}
Three datasets were analyzed. Dataset~(i) is the delay series under
0.25~eV excitation used for Figs.~\ref{fig:fig1}e and \ref{fig:fig3}. It
consists of 35 runs taken in two consecutive blocks. In each run the delay was
stepped once through a fixed list, and one spectrum was recorded at every
delay: Set~1 (25 runs) at $t = -15$, $-10$, 0, 2, 4, 10, 50 and 200~ps, and
Set~2 (10 runs) at $t = -15$, 6, 15 and 25~ps. Dataset~(i) thus contains 240
spectra, and each time delay point in Fig.~\ref{fig:fig3} is an average over
the runs that contain that delay. Dataset~(ii) consists of runs that scan the
delay across time zero under 1.20~eV and 0.25~eV excitation at an incident photon energy of 858.4 eV
(Fig.~\ref{fig:fig2}a), and dataset~(iii) of an incident-energy scan at
$t = 4$ and $-15$~ps (Fig.~\ref{fig:fig2}b,c). Each run contains its own
pre-pump reference spectrum $I_r^{\mathrm{ref}}(E)$: the $-15$~ps spectrum for
datasets (i) and (iii), and the average over $-5 < t < 0$~ps for
dataset~(ii). Spectra were normalized to the incident X-ray intensity, and the
energy transfer is defined as
$E_{\mathrm{loss}} = E_{\mathrm{in}} - E_{\mathrm{out}}$, so that
$E_{\mathrm{loss}} > 0$ ($< 0$) corresponds to Stokes (anti-Stokes)
scattering.

\paragraph{Delays and time resolution.}
The nominal delays of dataset~(i) were set by the time zero of the InGaAs scan
made just before it. The X-ray arrival time can drift after this scan; the
beamline phase cavity recorded it at 1~Hz throughout the measurement. For every
spectrum we averaged the phase-cavity readings taken while it was recorded and
subtracted the mean reading during the time-zero scan. The phase-cavity reading
falls when the X-rays arrive later, so the difference between the true and the
nominal delay of a spectrum is this change with its sign reversed. A positive
difference means that the X-rays arrived later than in the time-zero scan and
the true delay is longer than nominal.

\textit{Timing selection.} A spectrum was removed if its mean phase-cavity
reading, or the 5th or 95th percentile of its readings, deviated from the median
over all 240 spectra by more than a delay-dependent tolerance: 0.3~ps at $t=0$,
5~ps at $t<0$, and a quarter of the spacing to the neighboring delay at $t>0$
(0.5~ps at 2-6~ps). This removed 16 of the 240 spectra, 10 of them at $t=0$.
All pump-induced changes of dataset~(i) use the remaining 224 spectra.

\textit{Effective delays.} Supplementary Fig.~\ref{fig:sfig_pcav} shows this
difference at four of the delays of Fig.~\ref{fig:fig3}c,e, as histograms of
the 1-s phase-cavity readings of all runs that enter each point. The X-ray
timing varies from run to run and within a run, so we plot each delay point at
its nominal delay plus the shift averaged over its runs (blue lines). These
shifts range from 0.22 to 0.38~ps. Fig.~\ref{fig:fig3}c,e and the relaxation
fit use these effective delays; delays quoted in the text are nominal.

\textit{Time resolution.} The time resolution of a delay point combines the
X-ray pulse (200~fs FWHM), the 0.25~eV pulse (at most 350~fs FWHM), the
spread between the runs averaged at that delay (standard deviation
0.08-0.19~ps) and the variation of the phase cavity while one spectrum was
recorded (half its 5-95\% range, 0.15-0.19~ps, i.e.\ $\sigma = 0.09$-0.11~ps).
The last two together set the width of the histograms in Supplementary
Fig.~\ref{fig:sfig_pcav} (standard deviation 0.13-0.18~ps). These four terms
give $\sigma = 0.21$-0.27~ps, or 0.50-0.64~ps FWHM as the total time
resolution.

\paragraph{RIXS spectral analysis.}
\textit{Energy alignment.} In dataset~(i), each run was aligned on its own
elastic line. The pre-pump spectrum of every run (Set~1: average of the $-15$
and $-10$~ps spectra; Set~2: the $-15$~ps spectrum) was fitted between $-0.06$
and $0.39$~eV with a Gaussian elastic line, Voigt profiles for 1M and 2M and a
linear background. All 35 runs were fitted together: the Gaussian width was
fixed to the 79~meV resolution, the 1M and 2M energies and Lorentzian widths
were common to all runs, and the elastic position $E_0$, the amplitudes and the
background were free for each run. All spectra of a run were then shifted to
put $E_0$ at zero energy loss and interpolated onto a common 19.55~meV grid.

\textit{Equilibrium spectrum.} Fig.~\ref{fig:fig1}e shows the average of the
aligned pre-pump spectra of six runs of Set~1, with the standard error across
these runs, and a fit of the same model to this average. The fit gives $E_{1\mathrm M}=95.5\pm2.7$~meV.

\textit{Pump-induced changes.} In every dataset, the pump-induced change was
computed run by run against the same run's reference,
$\Delta I_r(E,t) = I_r(E,t) - I_r^{\mathrm{ref}}(E)$, and averaged across runs
only after subtraction; error bars are the standard error across runs. In
dataset~(i), each run's difference spectrum was integrated over fixed windows
before averaging: $-140$ to $-75$~meV (1M) and $-260$ to $-160$~meV (2M) on
the anti-Stokes side (Fig.~\ref{fig:fig3}e), and 0.25 to 0.75~eV on the Stokes
side (Fig.~\ref{fig:fig3}c, top). The $-10$~ps spectrum serves as a pre-pump
control. In dataset~(ii), $I_r(E,t)$ is the average over $0 < t < 5$~ps, and
the 1.20~eV and 0.25~eV scans were processed separately. In
dataset~(iii), the subtraction was done at each incident photon energy at
$t = 4$~ps, and the anti-Stokes weight
$\Delta W(h\nu) = \int_{-0.4~\mathrm{eV}}^{0} \Delta I(E; h\nu)\,dE$
(Fig.~\ref{fig:fig2}c) was compared with the Ni $L_3$-edge absorption spectrum.

\textit{1M line width.} The 1M width was fitted on the Stokes side ($-0.04$ to
$0.45$~eV). At each delay, the spectra of all runs that contain it were fitted
together. Each spectrum has its own elastic line, 1M and 2M amplitudes and
linear background, while all of them share one 1M energy and one Lorentzian
width. The resolution is fixed to 79~meV and the 2M line shape to that of a fit
to the averaged pre-pump spectrum of all 35 runs. The reported width is the
FWHM of the resulting 1M Voigt profile. The reference width before the pump is
obtained in the same way from the pre-pump spectra of the same runs ($-15$ and
$-10$~ps for Set~1, $-15$~ps for Set~2). Uncertainties come from resampling
runs\cite{refDavison1997,refField2007}: in each of 1000 draws, runs were picked
at random with replacement within each set, and every delay was refitted
together with its reference from the same picked runs. The quoted interval
contains the central 95\% of the draws.

\textit{Relaxation fit.} The 2M trace from its maximum at 4~ps to 200~ps was
fitted at the effective delays with
$I_{2\mathrm{M}}(t) = A\exp[-(t - 4~\mathrm{ps})/\tau]$, weighted by the
standard errors and with the long-delay offset fixed to zero, giving
$\tau_{2\mathrm{M}} = 28 \pm 11$~ps ($1\sigma$, from the fit covariance).

\textit{Error bars in Fig.~\ref{fig:fig3}.} Error bars are standard errors
across runs, and the 95\% resampling interval for the 1M width. The grey peak
at $t=0$ in Fig.~\ref{fig:fig3}c,e and the rise of the dashed guides at $t=0$
both use the 0.50~ps FWHM time resolution of the 0~ps point.

\paragraph{First-principles magnon dispersion.}
The magnetic spectrum in Fig.~\ref{fig:fig1}b was obtained from
first-principles exchange parameters. Electronic-structure calculations were
performed using VASP\cite{refKresse1996DFT} with a hybrid
functional\cite{refHeyd2003DFT}. The resulting electronic structure was mapped
onto a Wannier-function Hamiltonian using Wannier90\cite{refPizzi2020DFT}, and
magnetic exchange parameters were evaluated using the magnetic force-theorem
approach implemented in TB2J\cite{refKorotin2015DFT,refHe2021DFT}. Rather than
truncating the magnetic Hamiltonian to a $J_1$-$J_2$ model, the full
calculated exchange network was retained in linear spin-wave theory. The
resulting three-dimensional one-magnon dispersion was used to construct the
bare two-magnon density of states at fixed total momentum along [100] from
$\Gamma$ to $X=(1,0,0)$. The exchange parameters and the definition of the
density of states are given in Supplementary Note~1.

\paragraph{Minimal-model calculations.}
Magnon-magnon interaction effects and the collision estimate use the
isotropic, collinear $J_1$-$J_2$ Heisenberg model with $S=1$,
$J_1=-1.37$~meV and $J_2=19.01$~meV\cite{refHutchings1972}. Momenta are given
in units of $2\pi/a$, with $a$ the cubic lattice constant. $\Qz$ lies inside
the magnetic Brillouin zone of the type-II structure, close to the maximum of
the one-magnon dispersion along [100] (Supplementary Note~1). The interacting
two-magnon response in Supplementary Fig.~\ref{fig:sfig1} includes repeated
final-state scattering from the quartic magnon-magnon interaction through a
Bethe-Salpeter calculation. It is an incoherent average of the responses to
the nine representative bond sources, with weights
$w_{\dd}=|J_{\dd}|/\sum_{\dd'}|J_{\dd'}|$. Supplementary Note~1
specifies the bond convention, source vertices, pair interaction kernel and
numerical parameters. Supplementary Note~2 gives the phonon-assisted injection
constraints and the dilute-density $2\rightarrow2$ collision estimate,
including its dependence on the assumed distribution, pump fluence and
excitation efficiency.

\section*{Data availability}

The data supporting the findings of this article are openly available in the Stanford Digital Repository at \url{https://doi.org/10.25740/nk931cs4375}.

\section*{Acknowledgements}

This work was supported as part of the Center for Energy Efficient Magnonics, an Energy Frontier Research Center funded by the U.S. Department of Energy, Office of Science, Basic Energy Sciences, under Award number DE-AC02-76SF00515. Use of the Linac Coherent Light Source (LCLS), SLAC National Accelerator Laboratory, is supported by the U.S. Department of Energy, Office of Science, Office of Basic Energy Sciences under Contract No. DE-AC02-76SF00515. Use of the Stanford Synchrotron Radiation Lightsource, SLAC National Accelerator Laboratory, is supported by the U.S. Department of Energy, Office of Science, Office of Basic Energy Sciences under Contract No. DE-AC02-76SF00515. We thank Cheng-Tai Kuo, Paul Corbae, and Jun-Sik Lee for assistance with the Ni \(L_{2,3}\)-edge X-ray absorption measurements at SSRL Beamline 13-3.

\section*{Author contributions}

X.P.Y., C.T.P., E.B., O.D., Y.L., L.S., D.J., J.J.T., G.L.D., W.-S.L. and A.H.R. performed RIXS measurements. M.C., Y.L., G.C., C.S. and M.C.H. prepared the laser pump for the time-resolved RIXS experiment. E.B. and H.Y.H. grew NiO thin films. X.P.Y. and C.-M.L. performed theoretical modelling. H.P. and D.P. performed first-principles calculations. X.P.Y. and A.H.R. performed the X-ray absorption measurement. X.P.Y. and A.H.R. interpreted and analysed the data. A.H.R. supervised the project. X.P.Y. and A.H.R. wrote the paper, with input from all co-authors.

\section*{Competing interests}
The authors declare no competing interests.

\singlespacing

\clearpage
\appendix
\setstretch{1.15}
\section*{Supplementary Information}
\renewcommand{\citeform}[1]{S#1}%
\pagestyle{plain}

\setcounter{figure}{0}
\renewcommand{\thefigure}{S\arabic{figure}}
\renewcommand{\theHfigure}{S\arabic{figure}}

\setcounter{table}{0}
\renewcommand{\thetable}{S\arabic{table}}
\renewcommand{\theHtable}{S\arabic{table}}

\setcounter{equation}{0}
\renewcommand{\theequation}{S\arabic{equation}}
\renewcommand{\theHequation}{S\arabic{equation}}

\section*{Supplementary Note 1: Two-magnon response of the minimal spin-wave model}

\subsection*{Spin-wave model}

We start from the Heisenberg Hamiltonian
\begin{equation}
H=\sum_{\RR}\sum_{\dd\in\Dplus}J_{\dd}\,\mathbf S_{\RR}\cdot\mathbf S_{\RR+\dd}
\label{eq:HJ_supp}
\end{equation}
with $S=1$, where $\Dplus$ contains each of the six
nearest-neighbor ($J_1=-1.37$~meV) and three next-nearest-neighbor
($J_2=19.01$~meV) bonds of the fcc Ni lattice, taken from inelastic neutron
scattering\cite{S:refHutchings1972}. In the type-II order, with ordering vector $\QQ_{\rm AF}=(\pi,\pi,\pi)$, the spins are
parallel within each (111) plane and antiparallel between neighboring planes.
A bond therefore connects parallel spins if it lies within a (111) plane and
antiparallel spins otherwise, which we write as $c_{\dd}=\cos(\QQ_{\rm AF}\cdot\dd)=+1$ or $-1$. Three of the six $J_1$ bonds lie within a (111) plane and are
parallel, $D_\parallel=\{(\tfrac12,-\tfrac12,0),(\tfrac12,0,-\tfrac12),
(0,\tfrac12,-\tfrac12)\}$; the other three $J_1$ bonds and all three $J_2$
bonds are antiparallel and form the set $D_{\rm AP}$. The measured momentum is $\Qz=(\pi/a,0,0)$, or $(\tfrac12,0,0)$ in units of
$2\pi/a$. It lies inside the magnetic Brillouin zone of the type-II
structure, whose boundary along [100] is at $(\tfrac34,0,0)$. $\Qz$ is the zone-boundary point of the
simple-cubic lattices formed by the dominant $J_2$ bonds, into which the
type-II structure decouples for $J_1\to0$\cite{S:refHutchings1972}. Because
$|J_1/J_2|\approx0.07$, the dispersion along [100] is nearly symmetric about
$\Qz$ and reaches its maximum close to it, which is why a previous RIXS study refers to
this point as the [001] zone boundary\cite{S:refBetto2017}.

After a local rotation that aligns every ordered moment with $+\tilde z$, we
apply the Holstein-Primakoff transformation\cite{S:refHP1940},
$\tilde S_i^z=S-a_i^\dagger a_i$ and
$\tilde S_i^+=\sqrt{2S-a_i^\dagger a_i}\;a_i
\simeq\sqrt{2S}\,(1-a_i^\dagger a_i/4S)\,a_i$.
Expanding in powers of these bosons gives
$H=H^{(0)}+H^{(2)}+H^{(4)}+\cdots$, where $H^{(0)}$ is the classical
ground-state energy, $H^{(2)}$ is the non-interacting magnon Hamiltonian and
$H^{(4)}$ is the four-magnon interaction. Three-magnon terms vanish for
collinear order\cite{S:refZhitomirsky2013,S:refChoe2025}. With
$a_{\RR}=N^{-1/2}\sum_{\kk}e^{i\kk\cdot\RR}a_{\kk}$, the quadratic term is
\begin{equation}
H^{(2)}=\frac12\sum_{\kk}\Phi_{\kk}^\dagger D(\kk)\,\Phi_{\kk},\qquad
\Phi_{\kk}=\begin{pmatrix}a_{\kk}\\ a_{-\kk}^\dagger\end{pmatrix},\qquad
D(\kk)=\begin{pmatrix}A_{\kk}&-B_{\kk}\\-B_{\kk}&A_{\kk}\end{pmatrix},
\label{eq:H2_supp}
\end{equation}
with
\begin{align}
A_{\kk}&=6SJ_2+2SJ_1\Big[\cos\tfrac{k_x-k_y}{2}+\cos\tfrac{k_x-k_z}{2}
+\cos\tfrac{k_y-k_z}{2}\Big],\nonumber\\
B_{\kk}&=2SJ_2\big(\cos k_x+\cos k_y+\cos k_z\big)
+2SJ_1\Big[\cos\tfrac{k_x+k_y}{2}+\cos\tfrac{k_x+k_z}{2}
+\cos\tfrac{k_y+k_z}{2}\Big].
\label{eq:AB_supp}
\end{align}
The paraunitary (Bogoliubov) transformation\cite{S:refColpa1978}
$a_{\kk}=u_{\kk}\alpha_{\kk}+v_{\kk}\alpha_{-\kk}^\dagger$, with
$u_{\kk}=[(A_{\kk}/\varepsilon_{\kk}+1)/2]^{1/2}$ and
$v_{\kk}=\operatorname{sgn}(B_{\kk})[(A_{\kk}/\varepsilon_{\kk}-1)/2]^{1/2}$,
diagonalizes $H^{(2)}$ with magnon energies
\begin{equation}
\varepsilon_{\kk}=\sqrt{A_{\kk}^2-B_{\kk}^2},
\label{eq:dispersion_supp}
\end{equation}
giving $\varepsilon_{\Qz}=105.6$~meV and $\varepsilon_{\kk\to0}=0$.

The four-magnon interaction is
\begin{align}
H^{(4)}=\frac1N\sum_{1234}\Big[&\delta_{\kk_1+\kk_2,\kk_3+\kk_4}\,V_{1234}\,
a_1^\dagger a_2^\dagger a_3a_4\nonumber\\
&+\delta_{\kk_1,\kk_2+\kk_3+\kk_4}\big(W_{1;234}\,a_1^\dagger a_2a_3a_4
+\mathrm{H.c.}\big)\Big],
\label{eq:H4_supp}
\end{align}
where the labels $1,\dots,4$ denote momenta $\kk_1,\dots,\kk_4$ and
$\delta_{\kk,\kk'}$ is the Kronecker delta, which equals 1 when $\kk$ and
$\kk'$ differ by a reciprocal-lattice vector and 0 otherwise, so that only
momentum-conserving processes contribute. The vertices are
\begin{align}
V_{1234}&=\sum_{\dd\in D_\parallel}J_{\dd}\Big[e^{i(\kk_4-\kk_2)\cdot\dd}
-\tfrac14\big(e^{i\kk_1\cdot\dd}+e^{-i\kk_2\cdot\dd}+e^{-i\kk_3\cdot\dd}
+e^{i\kk_4\cdot\dd}\big)\Big]
-\sum_{\dd\in D_{\rm AP}}J_{\dd}\,e^{i(\kk_4-\kk_2)\cdot\dd},\nonumber\\
W_{1;234}&=\frac14\sum_{\dd\in D_{\rm AP}}J_{\dd}
\big(e^{-i\kk_3\cdot\dd}+e^{i\kk_4\cdot\dd}\big).
\label{eq:VW_supp}
\end{align}
The $V$ term scatters two Holstein-Primakoff bosons and conserves their total
number. Its contribution $e^{i(\kk_4-\kk_2)\cdot\dd}$ is the density-density interaction $c_{\dd}J_{\dd}n_in_j$ between
Holstein-Primakoff bosons on neighboring sites, which is dominated by the
$J_2$ bonds; the
remaining terms on parallel bonds are Holstein-Primakoff corrections to magnon
hopping. The $W$ term arises from Holstein-Primakoff corrections to pair
creation and annihilation on antiparallel bonds and changes the boson number
by two.

\subsection*{Interacting two-magnon response}

We follow the interacting spin-wave calculation of two-magnon spectra by Elliott
and Thorpe\cite{S:refElliottThorpe1969} and by Lorenzana and
Sawatzky\cite{S:ref6,S:ref7}. The calculation has two ingredients: how a magnon
pair is created and how the two magnons interact.

\textit{Pair creation.} In phonon-assisted absorption, the light and the phonon
perturb the exchange coupling between two neighboring spins, which creates a
pair of magnons\cite{S:ref6,S:ref7}. We describe this process, and the
$\Delta M_S=0$ part of the RIXS response, with one operator for each bond
direction $\dd$,
\begin{equation}
O_{\dd}(\PP)=\sum_{\RR}e^{i\PP\cdot(\RR+\dd/2)}\,
\mathbf S_{\RR}\cdot\mathbf S_{\RR+\dd},
\end{equation}
which is the exchange energy of all bonds along $\dd$, summed with the phase of
the momentum transfer $\PP$. The phase factor makes the created magnon pair carry the momentum $\PP$
transferred by the probe, as in the two-magnon operators used for indirect
RIXS \cite{S:refvandenBrink2007,S:refNagao2007,S:refForte2008}. Because $O_{\dd}$ conserves the total $S^z$, it
creates only $\Delta M_S=0$ bimagnons; the single-site $\Delta M_S=\pm2$ double
magnon\cite{S:ref11} is not included. In magnon operators, the part of $O_{\dd}$
that creates two magnons from the ground state is
$S\sum_{\kk}m_{\dd}(\kk,\PP)\,\alpha^\dagger_{\kk}\alpha^\dagger_{\PP-\kk}$,
where $m_{\dd}$ is the amplitude for creating the pair $(\kk,\PP-\kk)$,
\begin{equation}
\begin{split}
m_{\dd}(\kk,\PP)={}&\frac{c_{\dd}-1}{2}\big(u_{\kk}u_{\PP-\kk}+v_{\kk}v_{\PP-\kk}\big)
\cos(\mathbf q\cdot\dd)\\
&+\Big[\frac{1+c_{\dd}}{2}\cos(\mathbf q\cdot\dd)-c_{\dd}\cos\frac{\PP\cdot\dd}{2}\Big]
\big(u_{\kk}v_{\PP-\kk}+v_{\kk}u_{\PP-\kk}\big),
\end{split}
\label{eq:mdelta_supp}
\end{equation}
with $\mathbf q=\kk-\PP/2$. Without interactions, the two-magnon spectrum is
the sum of $|m_{\dd}|^2$ over all pairs with energy
$\varepsilon_{\kk}+\varepsilon_{\PP-\kk}=E$.

\textit{Interaction between the two magnons.} The two magnons are created at the
two ends of a bond and interact through $H^{(4)}$. A two-magnon state with
total momentum $\PP$ is a superposition
$\sum_{\kk}\psi(\kk)\,\alpha^\dagger_{\kk}\alpha^\dagger_{\PP-\kk}|0\rangle$,
where the pair wavefunction $\psi(\kk)$ is the amplitude for one magnon to have
momentum $\kk$ and the other $\PP-\kk$. Its energy $E$ follows from the
two-magnon Schr\"odinger equation\cite{S:refWortis1963,S:refElliottThorpe1969,S:ref4}
\begin{equation}
\big[E-\varepsilon_{\kk}-\varepsilon_{\PP-\kk}\big]\psi(\kk)
=\frac{1}{2N}\sum_{\kk'}V_{\PP}(\kk,\kk')\,\psi(\kk'),
\label{eq:pair_schrodinger_supp}
\end{equation}
where $V_{\PP}(\kk,\kk')$, the interaction kernel, is the amplitude for
scattering the pair $(\kk',\PP-\kk')$ into $(\kk,\PP-\kk)$. Without
interaction the right-hand side vanishes and each pair has the free energy
$\varepsilon_{\kk}+\varepsilon_{\PP-\kk}$. The dominant part of $V_{\PP}$ is the density-density
term $c_{\dd}J_{\dd}n_in_j$ of Eq.~\eqref{eq:VW_supp}: compared with two
separated magnons, two magnons on the two ends of bond $\dd$ have an extra energy
$c_{\dd}J_{\dd}$, which lowers the pair energy by $J_2$ on the dominant $J_2$
bonds. The other terms are smaller because they are either proportional to
$J_1$ or require a Bogoliubov factor $v$, which is typically 0.15 for the magnons
that form the two-magnon peak. Keeping only the density-density term and setting
$u=1$, $v=0$ gives
\begin{equation}
V_{\PP}(\kk,\kk')\simeq4\sum_{\dd\in\Dplus}c_{\dd}J_{\dd}\,
\cos(\mathbf q\cdot\dd)\cos(\mathbf q'\cdot\dd),
\qquad \mathbf q'=\kk'-\PP/2,
\label{eq:Vapprox_supp}
\end{equation}
where $\cos(\mathbf q\cdot\dd)$ measures how strongly a pair with relative
momentum $\mathbf q$ places its two magnons on the two ends of bond $\dd$. This
term already gives the
peak two-magnon energies within 2~meV of the full calculation, which
uses the full $V_{\PP}$ that contains all terms of Eq.~\eqref{eq:VW_supp}
after the Bogoliubov transformation. One-in/three-out processes, which cause
magnon decay\cite{S:refZhitomirsky2013}, are omitted.

\textit{Numerical details.} The spectrum in Supplementary
Fig.~\ref{fig:sfig1} averages the nine bond spectra with weights
$|J_{\dd}|/\sum_{\dd'}|J_{\dd'}|$, which assumes that the excitation strength of
each bond scales with its exchange strength. Momentum sums use 8192 scrambled Sobol
points\cite{S:refSobol1967} and a Lorentzian broadening of 2~meV. 

\subsection*{First-principles comparison}

The spectrum in Fig.~\ref{fig:fig1}b uses exchange interactions calculated
with VASP\cite{S:refKresse1996DFT} and a hybrid functional\cite{S:refHeyd2003DFT},
mapped onto a Wannier-function Hamiltonian with Wannier90\cite{S:refPizzi2020DFT}
and evaluated with the magnetic force-theorem approach in
TB2J\cite{S:refKorotin2015DFT,S:refHe2021DFT}. In the convention of
Eq.~\eqref{eq:HJ_supp}, the dominant interaction is
$J_2^{\rm DFT}=16.99$~meV, close to the neutron value, while the
nearest-neighbor exchanges are 0.1-0.3~meV. Retaining the full calculated
exchange network in linear spin-wave theory, we compute the bare two-magnon
density of states\cite{S:refvandenBrink2007,S:refForte2008}
$D(\PP,E)=N_{\kk}^{-1}\sum_{\kk,\nu,\mu}
\delta[E-\omega_\nu(\kk)-\omega_\mu(\PP-\kk)]$ (Fig.~\ref{fig:fig1}b). The minimal $J_1$-$J_2$ model above is used mainly for the collision estimate in
Supplementary Note~2.

\section*{Supplementary Note 2: Magnon kinetics after phonon-assisted excitation}

This note estimates the magnon density created by the 0.25~eV pump and the
time between magnon-magnon collisions in the minimal model of Supplementary
Note~1.

\subsection*{Pumped magnon density}

In phonon-assisted absorption, a pump photon creates a magnon pair with total
momentum $\mathbf P_{2M}$ and energy $E_{2M}$, together with a phonon of branch $\nu$ and
momentum $\vq$:
\begin{equation}
\hbar\omega_{\rm pump}=E_{2M}+\hbar\Omega_{\vq\nu},
\qquad
\mathbf P_{2M}+\vq\simeq0,
\label{eq:pump_selection_supp}
\end{equation}
where the photon momentum is negligible. Early analyses of the NiO
absorption sideband considered zone-center optical phonons\cite{S:ref1}, which
create pairs with $\mathbf P_{2M}\simeq0$. Later phonon-assisted multi-magnon theory allows
phonons from the whole Brillouin zone, with a strongly momentum-dependent
matrix element\cite{S:ref6,S:ref7}, so pairs can also be created at finite $\mathbf P_{2M}$.

Let $\eta$ be the fraction of absorbed pump photons that each create one
magnon pair. For a film of thickness $d$ and absorption coefficient $\alpha$,
the number of magnons per Ni is
\begin{equation}
\bar n_{\rm mag}
=\frac{2\eta F_{\rm inc}\left(1-e^{-\alpha d}\right)}
{n_{\rm Ni}\,d\,\hbar\omega_{\rm pump}}
\simeq\frac{2\eta\,\alpha F_{\rm inc}}{n_{\rm Ni}\hbar\omega_{\rm pump}}
\qquad(\alpha d\ll1),
\label{eq:nactualfluence_supp}
\end{equation}
where $F_{\rm inc}$ is the incident pump fluence (energy per unit area per
pulse), $F_{\rm inc}(1-e^{-\alpha d})$ is the fluence absorbed in the film,
and $n_{\rm Ni}=4/a^3=5.5\times10^{22}$~cm$^{-3}$ is the Ni density
($a=4.18$~\AA). Reflection losses are neglected. At
$\hbar\omega_{\rm pump}=0.25$~eV the absorption coefficient is
$\alpha\simeq40$~cm$^{-1}$\cite{S:ref8}, so $\alpha d\simeq5\times10^{-4}$ for
our 120~nm film and $\bar n_{\rm mag}$ does not depend on $d$.
Equation~\eqref{eq:nactualfluence_supp} then gives $\bar n_{\rm mag}\simeq1.24\times10^{-3}\,\eta$ at the experimental fluence
$F_{\rm inc}=34$~mJ\,cm$^{-2}$.

The same expression gives the energy absorbed from the 1.20~eV control pulse, for which we estimate $\alpha\simeq270$~cm$^{-1}$ from
the low-temperature absorption spectrum reported by Newman and
Chrenko\cite{S:ref8}. For $d=120$~nm the absorbed fractions $1-e^{-\alpha d}$ are
$4.8\times10^{-4}$ at 0.25~eV and $3.2\times10^{-3}$ at 1.20~eV. The incident
fluences of 34 and 17.9~mJ\,cm$^{-2}$ therefore correspond to absorbed fluences
of about 16 and 58~$\mu$J\,cm$^{-2}$, so the 1.20~eV pulse deposits about 3.5
times more energy in the film. The 0.25~eV pump nevertheless produces the
larger anti-Stokes response, consistent with resonant excitation of the
phonon-assisted two-magnon absorption.

\subsection*{Magnon-magnon collision time}

The number-conserving part of $H^{(4)}$ (Eq.~\eqref{eq:H4_supp}) scatters two
magnons, $\kk_1+\kk_2\to\kk_3+\kk_4$, conserving crystal momentum and energy.
Writing $n_i$ and $\varepsilon_i$ for the occupation and energy of the magnon
mode $\kk_i$, these collisions change the occupation of mode
$\kk_1$ at the rate\cite{S:refBayrakci2013,S:refChoe2025,S:refHua2024}
\begin{align}
\dot n_1
&=\frac{2\pi}{\hbar}\,\frac{1}{2N^2}\sum_{\kk_2\kk_3\kk_4}
|V_{12\to34}|^2\,
\delta_{\kk_1+\kk_2,\,\kk_3+\kk_4}\,
\delta(\varepsilon_1+\varepsilon_2-\varepsilon_3-\varepsilon_4)
\nonumber\\
&\quad\times\left[(1+n_1)(1+n_2)\,n_3n_4-n_1n_2\,(1+n_3)(1+n_4)\right],
\label{eq:Boltzmann_supp}
\end{align}
where $N$ is the number of Ni sites and the factor $1/2$ compensates for
counting each final state twice in the sum over $\kk_3$ and $\kk_4$.
The vertex $V_{12\to34}$ is the two-in/two-out vertex of the Bogoliubov magnons
$\alpha_{\kk}$ derived from Eq.~\eqref{eq:H4_supp}. At fixed
$\mathbf P_{2M}=\kk_1+\kk_2$ it is the pair-scattering kernel of the ladder
calculation in Supplementary Note~1.

The pumped state contains many magnon pairs, and a
collision can involve one magnon from each of two different pairs. Consider
the pairs $(\kk,-\kk)$ and $(\kk',-\kk')$, both with $\mathbf P_{2M}=0$. If $\kk$ and
$\kk'$ scatter to $\kk+\mathbf p$ and $\kk'-\mathbf p$, the magnons at $-\kk$
and $-\kk'$ are unaffected. The total momentum of the four magnons is
unchanged, but the magnon at $\kk+\mathbf p$ and its former partner at $-\kk$
now have total momentum $\mathbf p$, and the other two have $-\mathbf p$. Such
collisions can therefore move weight from pairs near $\mathbf P_{2M}=0$ to pairs at
finite $\mathbf P_{2M}$, including $\Qz$, which a single-pair calculation cannot
describe. 

In the dilute limit, $n_i\ll1$, the loss term of Eq.~\eqref{eq:Boltzmann_supp}
gives the rate at which a pumped magnon collides with the others. Writing
$n_{\kk}=\bar n_{\rm mag} f(\kk)$ with $N^{-1}\sum_{\kk}f(\kk)=1$, and averaging the
colliding magnon over the same distribution, gives
\begin{equation}
\frac{1}{\tau_{2\rightarrow2}}=K\bar n_{\rm mag},
\qquad
K=\frac{\pi}{\hbar}\left\langle|V_{12\to34}|^2\,
\delta(\varepsilon_1+\varepsilon_2-\varepsilon_3-\varepsilon_4)\right\rangle,
\label{eq:taun_supp}
\end{equation}
where the average is over $\kk_1$ and $\kk_2$ drawn independently from
$f(\kk)$ and $\kk_3$ uniform in the Brillouin zone, with
$\kk_4=\kk_1+\kk_2-\kk_3$. The coefficient $K$ therefore depends on $f(\kk)$ and the
$H^{(4)}$ vertex.

\begin{figure}[t]
    \centering
    \includegraphics[width=\linewidth]{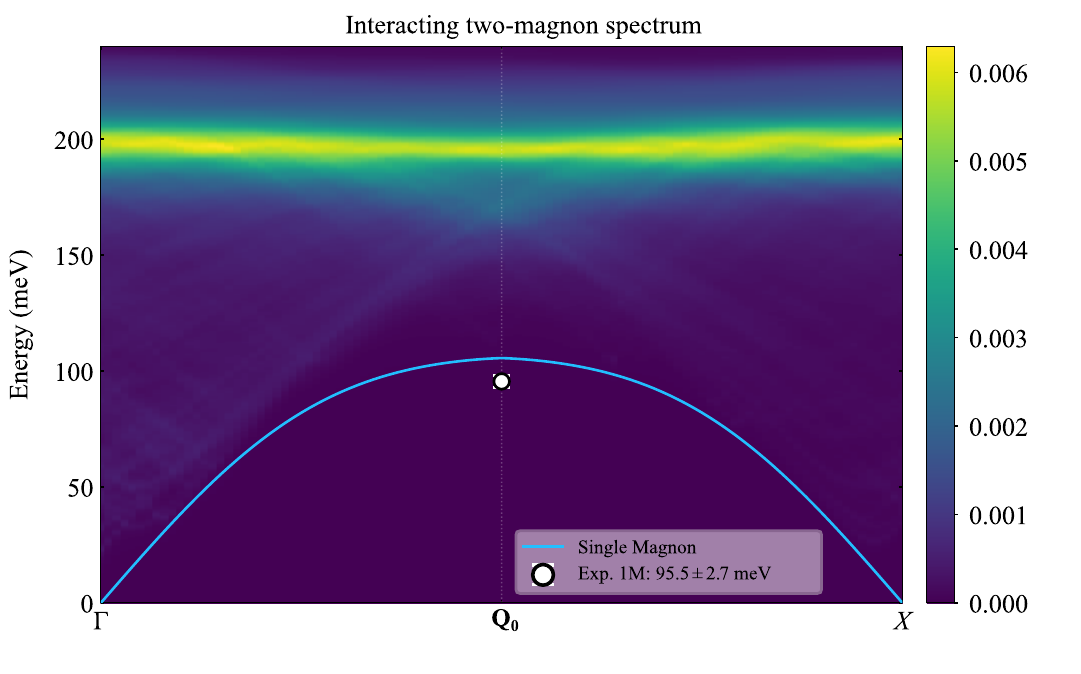}
    \caption{
\textbf{Two-magnon response of the minimal $J_1$-$J_2$ model.}
Interacting two-magnon ($\Delta M_S=0$) response along [100] from $\Gamma$
through $\Qz$ to $X=(1,0,0)$, including repeated final-state scattering from the
quartic magnon-magnon interaction (Supplementary Note~1). The color scale shows
the $|J_{\dd}|$-weighted incoherent average of the bond-resolved responses in
arbitrary units, without normalization. The cyan curve is the one-magnon
dispersion of the same model, and the white circle marks the experimental
one-magnon energy at $\Qz$, $95.5\pm2.7$~meV.
}
    \label{fig:sfig1}
\end{figure}

To construct $f(\kk)$ we keep pairs $(\kk,\mathbf P_{2M}-\kk)$ whose energy
$\varepsilon_{\kk}+\varepsilon_{\mathbf P_{2M}-\kk}$ lies within $198\pm20$~meV. As a
proxy, each pair is weighted by
$W(\mathbf P_{2M},\kk)=\sum_{\dd}w_{\dd}|m_{\dd}(\kk,\mathbf P_{2M})|^2$, using the bond source
$m_{\dd}$ of Eq.~\eqref{eq:mdelta_supp} and the weights
$w_{\dd}=|J_{\dd}|/\sum_{\dd'}|J_{\dd'}|$ used for Supplementary Fig.~\ref{fig:sfig1}. Each pair contributes its weight at both constituent momenta,
\begin{equation}
f(\kk)\propto\sum_{\mathbf P_{2M},\kk'}W(\mathbf P_{2M},\kk')
\left[\delta_{\kk,\kk'}+\delta_{\kk,\mathbf P_{2M}-\kk'}\right],
\end{equation}
normalized so that $N^{-1}\sum_{\kk}f(\kk)=1$. We consider three limits for the pair momentum: (i) $\mathbf P_{2M}=\Gamma$,
corresponding to zone-center phonons; (ii) $\mathbf P_{2M}=\Qz$; and (iii) $\mathbf P_{2M}$ distributed
uniformly over the Brillouin zone before the energy window and $W$ are
applied.

We sample $2^{19}$ pairs $(\mathbf P_{2M},\kk)$ with a scrambled Sobol
sequence\cite{S:refSobol1967} and evaluate Eq.~\eqref{eq:taun_supp} by Monte
Carlo method, using four independent sets of $2^{16}$ collisions and replacing the
energy $\delta$-function by a normalized Gaussian of width $\sigma_E$. At $\sigma_E=3$~meV we obtain
$K=380$, 418 and 410~ps$^{-1}$ for cases (i)-(iii). Varying $\sigma_E$ from 1 to 5~meV changes these values by less than
4\%, and all values lie between 375 and 432~ps$^{-1}$. We take the
uniform $\mathbf P_{2M}$ value, $K=410$~ps$^{-1}$, as representative of a broad
distribution.

At the experimental fluence level,
Eq.~\eqref{eq:taun_supp} gives $\tau_{2\rightarrow2}\simeq1.97~{\rm ps}/\eta$, or
1.87-2.15~ps$/\eta$ over the full range of $K$. For order-unity conversion efficiency, a pumped magnon
therefore collides with another within a few picoseconds, comparable to the
4~ps at which the 2M response is largest. 

\subsection*{Magnon-phonon relaxation}

Ionic displacements $u_s$, where $s$ labels the ion and Cartesian direction,
modulate the exchange constants and add
$\delta H=\sum_{\langle ij\rangle}\sum_s(\partial J_{ij}/\partial u_s)\,u_s\,
\mathbf S_i\cdot\mathbf S_j$ to the Hamiltonian\cite{S:refStreib2019,S:refMaldonado2019}.
Expanding $u_s$ in phonon operators $b_{\vq\nu}$, which annihilate a phonon of
branch $\nu$ and momentum $\vq$, and $\mathbf S_i\cdot\mathbf S_j$ in the
Bogoliubov magnon operators $\alpha_{\kk}$ of Supplementary Note~1 gives,
among other terms, the number-conserving coupling
\begin{equation}
H_{\rm m\text{-}ph}=\sum_{\kk\vq\nu}g^{\nu}_{\kk\vq}\,
\alpha^\dagger_{\kk+\vq}\alpha_{\kk}
\left(b_{\vq\nu}+b^\dagger_{-\vq\nu}\right),
\label{eq:mph_supp}
\end{equation}
in which a magnon absorbs or emits one phonon with coupling constant
$g^{\nu}_{\kk\vq}$. This process transfers magnetic
energy to the lattice. It also changes the momentum of a pair: when one magnon
of a pair whose total momentum is at $\Qz$ scatters off a phonon, the pair
momentum changes by the phonon momentum. Magnon-phonon scattering therefore also
removes weight from the 2M response at $\Qz$.

\clearpage

\singlespacing

\clearpage
\singlespacing

\clearpage
\singlespacing
\noindent\makebox[\textwidth][c]{\includegraphics[width=\textwidth]{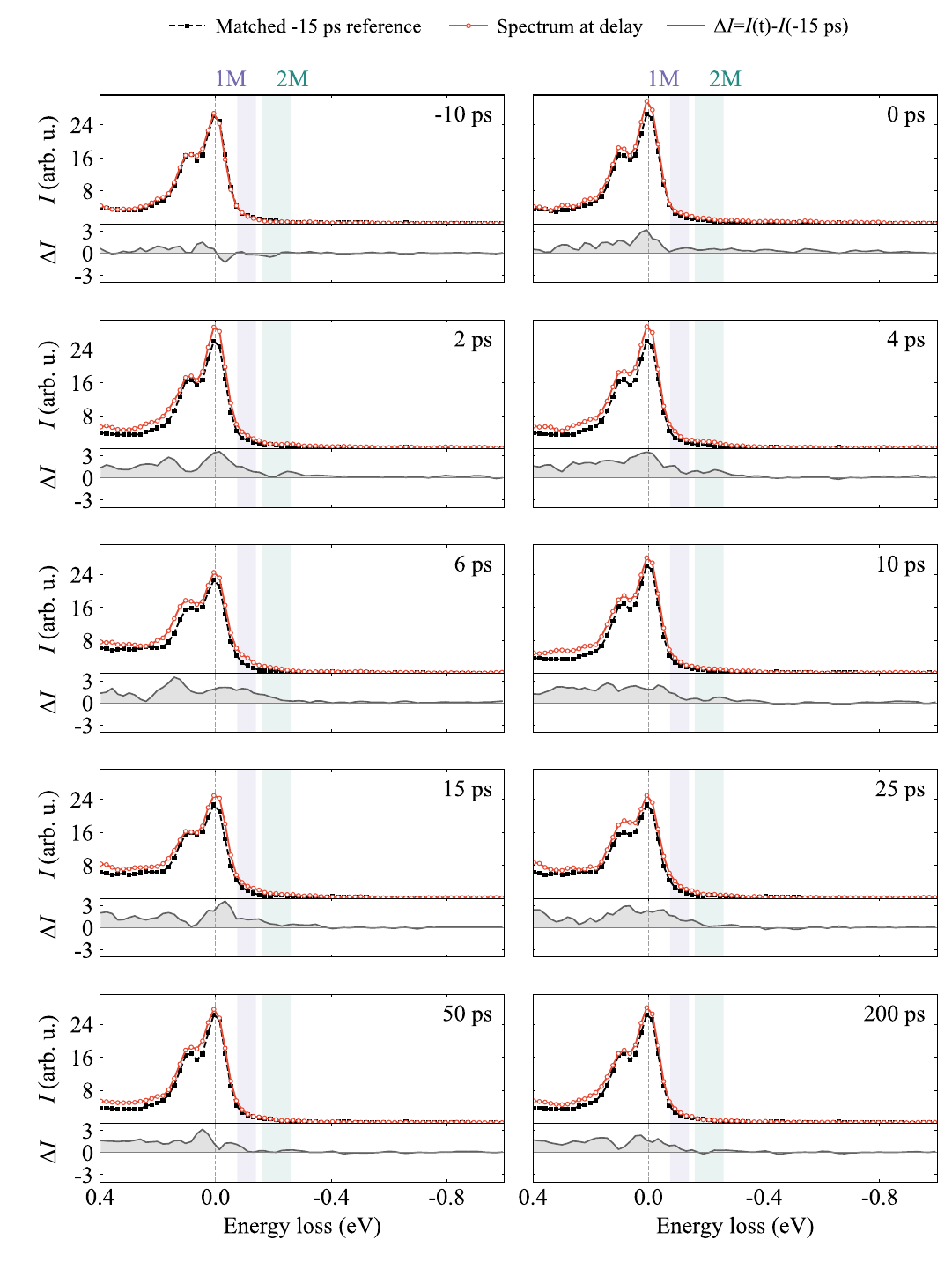}}

\clearpage
\noindent
\begin{minipage}{\textwidth}
\captionof{figure}{%
    \textbf{Complete time evolution of the pump-induced RIXS response.}
    RIXS spectra at all measured pump-probe delays, shown together with the
    corresponding run-matched $-15$~ps reference spectra.
    For each delay, the upper panel compares the measured spectrum
    (red circles) with its matched pre-pump reference
    (black dashed line and squares), while the lower panel shows the difference spectrum relative to the same run's reference,
    $\Delta I(E,t)=I(E,t)-I(E,-15~\mathrm{ps})$.
    The $-10$~ps measurement provides an independent pre-pump control.
    The shaded regions indicate the one-magnon (1M) and two-magnon (2M)
    energy windows used in the analysis.
    The extended energy-gain range is shown to test for additional pump-induced
    features outside the magnetic windows, including the charge-transfer-exciton
    response reported previously under above-bandgap excitation of NiO. No analogous
    feature is observed under the present mid-infrared pumping conditions.
}
\label{fig:sfig3}
\end{minipage}

\clearpage
\singlespacing

\begin{figure*}[t]
    \centering
    \includegraphics[width=0.5\textwidth]{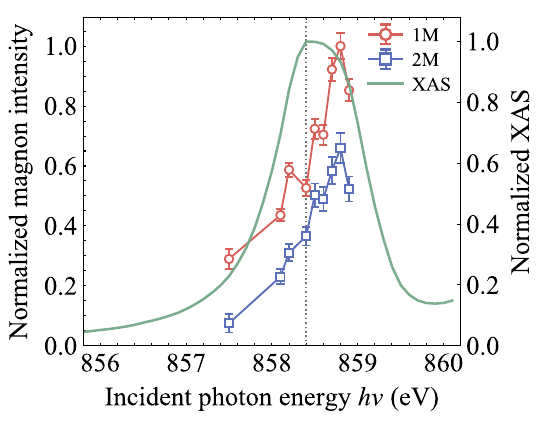}

    \caption{
\textbf{Incident-energy dependence of the Stokes-side one- and two-magnon responses across the Ni $L_3$ edge.}
Normalized integrated spectral weights of the one-magnon (1M) and
two-magnon (2M) components as a function of incident photon energy.
The magnetic spectral weights are extracted from fits to the equilibrium
RIXS spectra and normalized using a common intensity scale, preserving
the relative spectral weights of the two magnetic channels.
Error bars denote $1\sigma$ uncertainties obtained from the covariance
matrices of the spectral fits.
The green curve shows the equilibrium Ni $L_3$-edge X-ray absorption
spectrum measured in total-electron-yield mode and normalized to the
incident X-ray flux.
The vertical dotted line marks the Ni $L_3$ absorption maximum at
858.4~eV.
}
    \label{fig:sfig5}
\end{figure*}

\begin{figure}[t]
    \centering
    \includegraphics[width=0.65\linewidth]{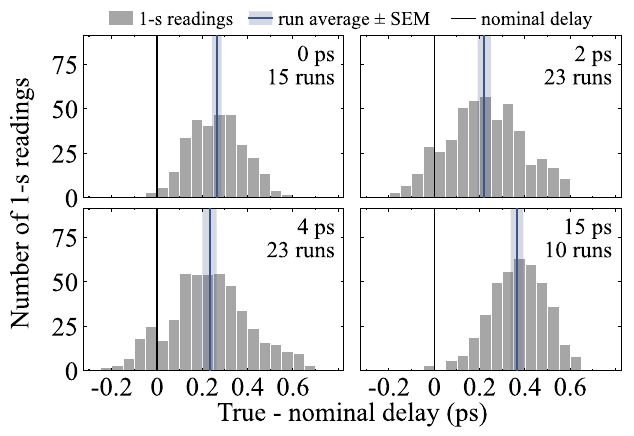}
    \caption{
\textbf{Phase-cavity correction of the delays in Fig.~\ref{fig:fig3}c,e.}
Difference between the true and the nominal delay at four of the time delays in
Fig.~\ref{fig:fig3}c,e, from the phase-cavity readings taken at 1~Hz while the
spectra were recorded, relative to the reading during the time-zero scan
(Methods). Each panel gives the nominal delay and the number of runs that enter
that point. The black line at zero is set by the X-ray timing
of the time-zero scan. Positive values mean that the X-rays arrived later, so
the true delay is longer than nominal. Grey bars are histograms of the 1-s
readings of these runs in 0.05~ps bins; readings taken during spectra removed
by the timing selection are left out. The blue line is the average over the
runs, and the shaded band is its standard error across runs. This average is
the shift applied to that point in Fig.~\ref{fig:fig3}c,e (0.22-0.37~ps here,
0.22-0.38~ps for all points), and its standard error (0.02-0.03~ps here,
0.02-0.04~ps for all points) is the uncertainty of the shift.
}
    \label{fig:sfig_pcav}
\end{figure}

\begin{figure}[t]
    \centering
    \includegraphics[width=\linewidth]{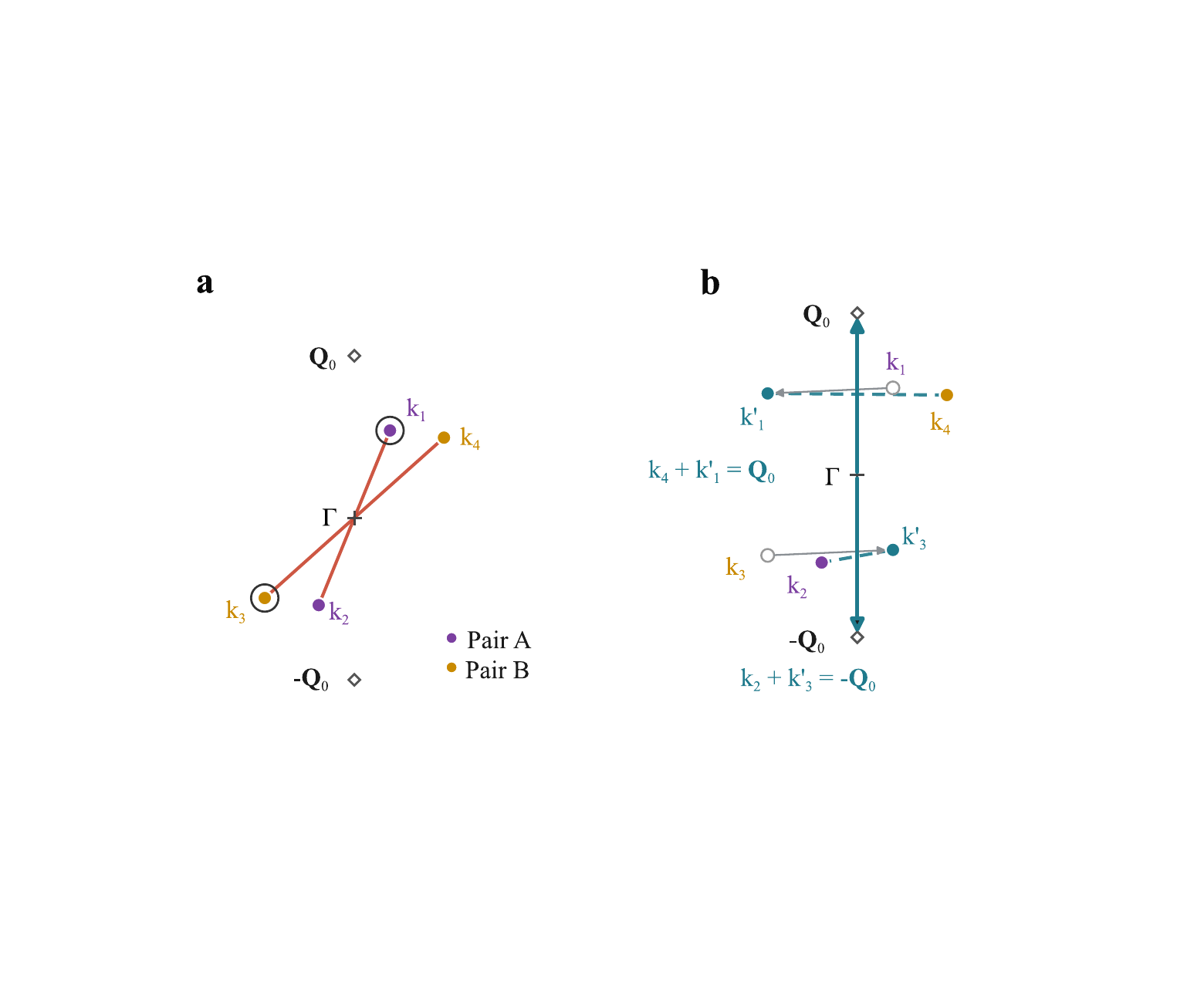}
    \caption{
    \textbf{Momentum-space picture of resonant two-magnon excitation and $H^{(4)}$ scattering.}
    \textbf{a,b,} A representative momentum- and energy-conserving
    $2\rightarrow2$ scattering event in the $k_y=0$ plane, with momenta
    expressed in units of $|\Qz|$.
    \textbf{a,} Two independently pumped magnon pairs,
    $A=(\mathbf{k}_1,\mathbf{k}_2)$ and
    $B=(\mathbf{k}_3,\mathbf{k}_4)$, are shown for the zone-center limiting
    case $P=0$, such that both pair chords pass through $\Gamma$.
    Circles indicate the two magnons participating in the collision, one
    drawn from each initial pair.
    \textbf{b,} The colliding magnons $\mathbf{k}_1$ and $\mathbf{k}_3$
    scatter to $\mathbf{k}'_1$ and $\mathbf{k}'_3$, while the spectator
    magnons $\mathbf{k}_2$ and $\mathbf{k}_4$ retain their momenta.
    Open circles indicate the pre-collision positions.
    Cross re-pairing of the outgoing many-magnon state gives
    $\mathbf{k}_4+\mathbf{k}'_1=+\Qz$ and
    $\mathbf{k}_2+\mathbf{k}'_3=-\Qz$.
    These possible outgoing pair configurations carry equal and
    opposite finite total momenta while conserving the total momentum of
    the full four-magnon state, $\sum \mathbf{P}=0$.
    Neither outgoing pair chord passes through $\Gamma$. 
    }
    \label{fig:sfig2}
\end{figure}

\clearpage
\singlespacing

\begin{figure}[p]
    \centering
    \includegraphics[width=\linewidth]{
        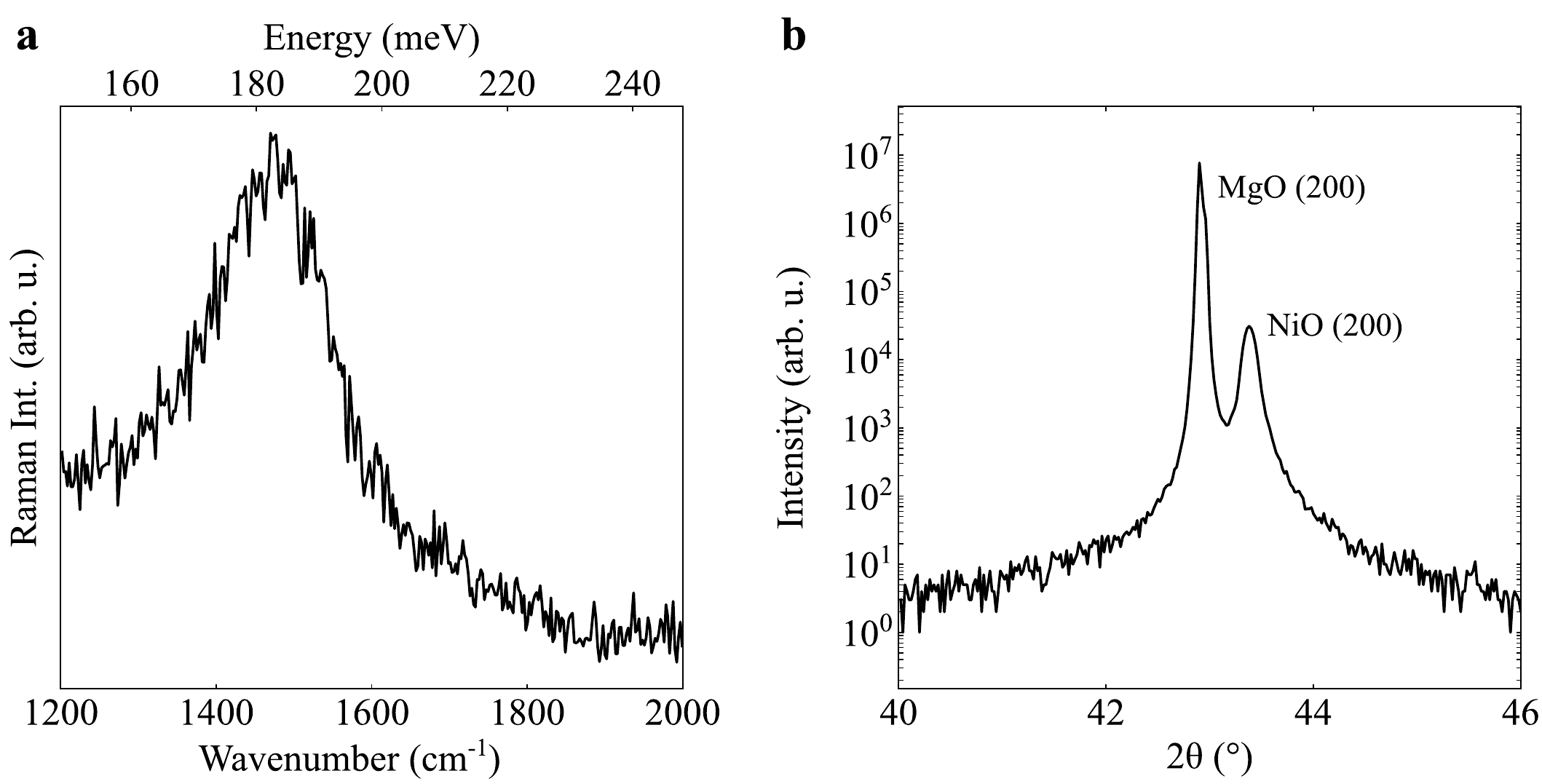
    }
    \caption{
    \textbf{Room-temperature Raman and X-ray diffraction characterization
    of the NiO film.}
    \textbf{a}, Raman spectrum of the approximately
    \qtyunit{120}{nm} thick NiO film grown on MgO (100),
    shown over 1200-2000~cm$^{-1}$.
    The two-magnon band is centered near 1475~cm$^{-1}$.
    \textbf{b}, XRD pattern of the film and substrate.
    The MgO (200) substrate and NiO (200) film reflections are labeled.
    The intensity is plotted on a logarithmic scale.
    Both measurements were performed at room temperature.
    }
    \label{fig:sfig_growth}
\end{figure}

\clearpage
\singlespacing
{\renewcommand{\refname}{Supplementary References}
\makeatletter\renewcommand{\@biblabel}[1]{[S#1]}\makeatother

\end{document}